\documentclass[manuscript,screen]{acmart}
\usepackage{hyperref}
\usepackage[ruled,vlined]{algorithm2e}
\usepackage{import}
\usepackage{tabularx}
\usepackage{bbm}
\usepackage{xcolor}
\usepackage{graphicx}
\usepackage{booktabs}
\usepackage{tikz}
\usepackage{environ}
\makeatletter
\newsavebox{\measure@tikzpicture}
\NewEnviron{scaletikzpicturetowidth}[1]{%
  \def\tikz@width{#1}%
  \def\tikzscale{1}\begin{lrbox}{\measure@tikzpicture}%
  \BODY
  \end{lrbox}%
  \pgfmathparse{#1/\wd\measure@tikzpicture}%
  \edef\tikzscale{\pgfmathresult}%
  \BODY
}
\makeatother

\usepackage{amsmath}
\usepackage{amssymb}
\newcommand{\R}{\mathbb{R}}
\usepackage{array}
\AtBeginDocument{%
  \providecommand\BibTeX{{%
    \normalfont B\kern-0.5em{\scshape i\kern-0.25em b}\kern-0.8em\TeX}}}

\setcopyright{acmcopyright}
\copyrightyear{2020}
\acmYear{2020}
\acmDOI{10.1145/1122445.1122456}

\makeatletter
\newcommand{\removelatexerror}{\let\@latex@error\@gobble}
\makeatother

\begin{document}

\title{Temporal State Machines: Using temporal memory to stitch time-based graph computations}

\author{Advait Madhavan}
\authornote{Both authors contributed equally to this research.}
\email{advait.madhavan@nist.gov}
\affiliation{%
  \institution{University of Maryland and National Institute of Standards and Technology}
}
\author{Matthew W. Daniels}
\email{matthew.daniels@nist.gov}

\authornotemark[1]

\author{Mark D. Stiles}
\email{mark.stiles@nist.gov}

\affiliation{%
  \institution{National Institute of Standards and Technology}
}


\begin{abstract}
%Race logic, an arrival-time-coded logic family, has demonstrated energy and performance improvements for a subset of applications ranging from dynamic programming to machine learning. In the absence of memory, or a mathematical framework for abstraction and composition, previous race logic demonstrations used \textit{ad hoc} mappings of algorithms into hardware. These restricted architectures, assembled from stateless, spatially arranged feed-forward paths, do not lend themselves to generalization. In order to systematize the development of temporally coded architectures, we associate race logic with the mathematical field tropical algebra. This association performs two functions. It provides a mapping between the mathematical primitives of tropical algebra and generalized race logic computations, guiding the design of temporally coded tropical circuits. It also serves as a useful framework for the expression of higher order timing-based algorithms. This abstraction, when combined with a natively temporal memory, makes it possible to partition feed-forward computations into stages and organize them into a state machine. We leverage analog memristor-based temporal memories to design a such a temporal state machine that operates purely on time-coded wavefronts. We implement a version of the well-studied Dijkstra's algorithm to evaluate this temporal state machine. This demonstration shows the promise of temporal computing to deliver significant energy and throughput advantages.
Race logic, an arrival-time-coded logic family, has demonstrated energy and performance improvements for applications ranging from dynamic programming to machine learning.  However, the \textit{ad hoc} mappings of algorithms into hardware result in custom architectures making them difficult to generalize. We systematize the development of race logic by associating it with the mathematical field called tropical algebra. This association between the mathematical primitives of tropical algebra and generalized race logic computations guides the design of temporally coded tropical circuits. It also serves as a framework for expressing high level timing-based algorithms. This abstraction, when combined with temporal memory, allows for the systematic generalization of race logic by making it possible to partition feed-forward computations into stages and organizing them into a state machine. We leverage analog memristor-based temporal memories to design a such a state machine that operates purely on time-coded wavefronts. We implement a version of Dijkstra's algorithm to evaluate this temporal state machine. This demonstration shows the promise of expanding the expressibility of temporal computing to enable it to deliver significant energy and throughput advantages.

\end{abstract}



\maketitle

\section{Introduction}
Energy efficiency is a key constraint when designing modern computers. The performance and efficiency of modern computers, which largely rely on Boolean encoding, can be attributed to developments across the computational stack from transistors through circuits, architectures, and other mid- to high-level abstractions. The recent stagnation of progress at the transistor level~\cite{hennessy2019new} is leading designers to make improvements at the lowest levels of the stack. These include re-imagining how data is encoded in physical states and introducing novel devices. The rationale is simple: making the fundamental mathematical operations required for computation more efficient can have a cascading effect on the whole architecture. However, novel encoding schemes and devices come with new trade-offs that differ from those of conventional Boolean computing schemes and which are not yet well understood.

%One approach to energy efficiency is analog computing, which takes advantage of analog data density and the ability of many analog devices to interface with external devices and signals. Another approach, which we pursue in the present paper, is known as unary computing. Unlike binary computing, which encodes information in base two, unary encodings use the unit radix, so that every symbol in a bitstring carries the same weight. Stochastic codes, for instance, produce ones or zeros in a bitstring according to some real-valued probability $p$. An arrival time code representing the value $t$, by constrast, switches from one symbol to the other precisely after the $t^\text{th}$ symbol in the string.
%Unary computing can operate in the digital domain and reap the benefits of traditional digital circuit techniques, while benefiting from the speed and energy advantages that may arise from alternative encodings. Recent examples of energy efficient unary computing schemes integrated with novel technologies can be found in Refs.~\cite{daniels2020energy, wu2020ugemm, madhavan2015race, najafi2018low,tzimpragos2020computational}.

%In this paper we focus on one specific unary computing scheme: 
In this paper, we focus on an arrival-time encoding known as race logic~\cite{madhavan2014race}. Since digital transitions (edges) account for much of the energy consumption in traditional computation, race logic encodes multi-bit information in a single edge per wire. The arrival time $t$ of this single edge is the value encoded by the signal. %A rising $0 \rightarrow 1$ electrical edge arriving at $t=2$ thus encodes neither zero nor one, but two. 
Encoding multiple bits on a single wire makes some operations very simple to implement. Standard \textsc{and} and \textsc{or} gates naturally implement the max and min functions; a unit delay element acts as an \textsc{increment} gate. A fourth logic gate, \textsc{inhibit}, allows its first (inhibiting) input to block the second input signal, if the inhibiting signal arrives earlier. %The operation of \textsc{and}, \textsc{or}, and \textsc{inhibit} are shown in Fig \ref{fig:1primitives}.

% \begin{figure}
%     \centering
%     \def\svgwidth{\textwidth}
%     \import{Figures/}{figure_1_revised.pdf_tex}
%     %\includegraphics[width=\linewidth]{Figures/1_primitives_app.pdf}
%     \caption{Race Logic primitives: The various panels show three out of the four main race logic primitives (top panels) as well as their behaviors (bottom panels).  The bottom panels show the voltage levels as a function of time for the input lines, A and B, and the output line, Q.  The information is encoded in the timing of the transition from low to high voltage. }
%     \label{fig:1primitives}
% \end{figure}

The development of race-logic-based architectures has been largely \textit{ad hoc}. Race logic was first developed to accelerate dynamic programming algorithms~\cite{madhavan2014race}, and its application space has expanded to include machine learning~\cite{RaceTrees} and sorting networks~\cite{najafi2018low}, demonstrating energy and performance advantages. Parallel development of logical frameworks~\cite{smith2018space, tzimpragos2020computational,am_isca_workshop2019}, novel device technologies~\cite{madhavan2020storing, vakili2020temporal}, and fabricated chips~\cite{madhavan20174} have contributed to a cross-stack effort to make this encoding scheme technologically viable. Here, we offer two important developments. 

The first development is a systematized method of building computing architectures.  An imporant step is to identify a suitable mathematical foundation that can express problems uniquely suited to a race logic approach. Formal logic, computation, and verification frameworks have been developed~\cite{smith2018space,tzimpragos2020computational,gtz_isca_workshop2019}. Continued progress requires identifying a mathematical algebra in which race logic \emph{algorithms} and \emph{state machines} are naturally expressed in the highly parallel dataflow contexts typical of temporal computing accelerators. We propose tropical algebra to be used in this context.

The second development is a compositional framework for programatically linking low-level subroutines into higher-order functions. This development expands race logic beyond one-shot application-specific circuits uses accelerator based architectures coupled with the absence of temporal memory technologies. Recent work has started to explore several device concepts for efficiently reading and writing time-coded signals~\cite{madhavan2020storing,vakili2020temporal}. The advantage of such memories is that they can directly interface with the temporal domain; read and write operations in such a memory can be performed without conversion to digital encoding. 

The introduction of temporal memory technologies allows race logic to serve as an efficient computational fabric for two distinct but compatible advances. First, because memory breaks symmetries related to translations of the time coordinate, a temporal computer equipped with a memory is no longer subject to the invariance constraint on time-coded functions outlined in Refs.~\cite{smith2018space, am_isca_workshop2019}. Lifting this restriction, allows tropical algebra to serve as a coherent algebraic context for designing and interfacing race logic circuits. Second, memories allow us to reach beyond specialized one-shot temporal computations. Primitive race logic operations can be composed and iterated upon by saving outputs of one circuit and rerunning that circuit on the saved state, the temporal equivalent of a classical state machine. In this paper, we develop the temporal state machine as a tool for accelerating tropical algebra in a generalizable computational fabric of high-efficiency race logic circuits.

Our contributions are:
\begin{itemize}
    \item A description of a temporal state machine that solves temporal problems in systematized parts, providing a clear computational abstraction for stitching larger computations out of primitive race logic elements. 
    \item An exposition of tropical algebra as a mathematical framework for working with temporal vectors. We explain the mapping into tropical algebra from race logic and how it provides a convenient mathematical setting for working with temporal computations.
    \item Augmentations to conventional 1T1R (1 transistor, 1 resistor) arrays that make the crossbar architecture natively perform fundamental tropical operations. We use this, and other temporal operations, to create a more general feed-forward temporal computation unit. 
    \item Demonstration and evaluation of a temporal state machine which uses Dijkstra's shortest path algorithm to find the minimal spanning tree on directed acyclic graphs. 
\end{itemize}{}

The paper is organized as follows. Section~\ref{sec:trop-race-isomorphism} briefly describes race logic and tropical algebra, showing the mapping between them. Based on that mapping, we describe circuit implementations of important tropical operations as the basic generators of higher order temporal functions. Section~\ref{sec:temporal-state-machine} introduces temporal state machines, explaining time-coded states and transition functions. We represent simple problems tropically and demonstrate how such a state machine can solve them in discrete steps. Section~\ref{sec:dijkstra} presents a case study implementating Dijkstra's algorithm on a temporal state machine, and proposes a purely temporal version of the algorithm. Performance and energy numbers follow in Section~\ref{sec:performance-results}, followed by a comparison with previous work and discussion in Section~\ref{sec:comp_tech_considerations}. 

\section{Tropical algebra and race logic: mapping between circuits and semirings}
\label{sec:trop-race-isomorphism}

\subsection{Race logic and temporal computing}
\label{subsec:rl_bckgnd}

Computing with time traces back to two communities, one, bio-inspired and the other purely efficiency oriented. The biological interest in precise timing relationships between spikes grew after the seminal works by Thorpe on the processing speed of the human visual system \cite{thorpe1990spike,guyonneau2005neurons} and on spike timing dependent plasticity by Bi and Poo \cite{bi1998synaptic}. From then, temporal wavefront computation \cite{ponulak2013rapid,izhikevich2006polychronization,vanrullen2005spike,ellender2010priming} in the biological community expanded to the machine learning and neuromorphic computing communities \cite{lagorce2015stick,orchard2015hfirst,osswald2017spiking}. References~\cite{tavanaei2018deep, wu2018spatio, bohte2000poutre, ponulak2010supervised, nair2020direct} show state of the art performance and learning strategies in temporal neural networks, while the neuromorphic computing community in Refs.~\cite{orchard2015hfirst,davidson2020spiking,khodagholy2015neurogrid, akopyan2015truenorth, furber2014spinnaker, davies2018loihi, sahay20202t}  developed hardware to emulate precise timing relationships in spiking neural activity. More recently, precise timing based codes in spiking neural networks perform a variety of applications such as graph processing~\cite{hamilton2019spike,kay2020neuromorphic}, median filtering~\cite{verzi2018computing}, image processing~\cite{verzi2018computing} and dynamic programming~\cite{aimone2019dynamic}. 
For several decades, the circuit community has independently been using time domain mixed signal analog techniques in Analog/Time to Digital Converters~\cite{yang2005time,oh2013time}, clock recovery circuits, phase and delay locked loops, phase detectors and arbiters. With shrinking voltage levels and diminishing headroom, the temporal domain becomes attractive for analog processing. With the interest in emerging computing paradigms, this community has contributed temporal coded complementary metal-oxide-semiconductor (CMOS) only computational approaches~\cite{miyashita2013ldpc,miyashita2017neuromorphic,dudek2006asynchronous,sayal201914,chen201919}.

Race logic sits between the aforementioned approaches in that it uses biologically-inspired wavefronts as the fundamental data structure, while using conventional digital CMOS circuits to compute. Race logic encodes information in the timing of rising digital edges and computes by manipulating delays between racing events. 
In the conventional Boolean domain, the electrical behaviour of wires changing voltage from ground to $V_{dd}$ is interpreted as changing from logic level 0 to logic level 1 at time $t$. In race logic, these wires are understood to encode each $t$ as their value, since the rising edge arrives at $t$ with respect to a temporal origin at $t=0$. 
In some cases, a voltage edge can fail to appear on a wire within the allotted operational time of a race logic computation. In these cases, we assign the value temporal infinity, represented by the $\infty$ symbol. 
%Each concrete numerical value for the arrival time of each rising edge is only meaningful relative to a reference time $t=0$. We will return to this crucial concept of a temporal origin throughout the paper, as it has wide ranging consequences on how race logic subcomputations can be arranged and composed. 

We define race logic without memory elements as \emph{pure} race logic, which accounts for most of the extant literature. We call race logic that uses dynamic memory elements \emph{stateful} or \emph{impure} race logic. Our goal here is to describe stateful race logic, but first we review issues that arise in pure race logic. The class of functions that can be implemented in pure race logic is constrained by physics~\cite{am_isca_workshop2019, smith2018space} through \emph{causality} and \emph{invariance}. The causal constraint, also called \emph{non-prescience}, requires that the output of a race logic function be greater than or equal to at least one of the function's inputs. Any output must be caused by an input that arrives either earlier than or simultaneously with that output. 

The invariance constraint arises because the circuit is indifferent to the choice of temporal origin. It is satisfied by race logic functions $f$ for which $f(t_1+\delta, t_2+\delta,\cdots,t_N+\delta) = f(t_1,t_2,\cdots,t_N)+\delta$; all operations in pure race logic must obey this equality. Invariance need not apply to impure circuits, which contain a memory or state element: such circuits perform differently at different times, depending on whether a memory element has been modified. From a programming perspective, a pure function is akin to a function in mathematics which always gives the same output when presented with the same input; an impure function is closer to a subroutine that can access and modify global variables.

%Pure race logic has been applied to a variety of graph- and tree-based problems in various different computing substrates from CMOS to novel emerging technologies. First presented in \cite{} as a solution to the DNA sequence alignment problem, it has been expanded to decision trees, machine learning algorithms, and sorting networks. The first experimental race logic chip was presented in \cite{}. The low activity factors of such computations result in large energy advantages leaving efficient spatial layout as a critical issue. These properties have been noted in similar temporally coded approaches as well.\cite{} 

\subsection{Tropical algebra}
\label{subsec:trop_bckgnd}

Named in honor of Brazilian mathematician Imre Simon, tropical algebra treats the tropical semiring $\mathbb{T}$.  In $\mathbb{T}$, the operations of addition and multiplication obey the familiar rules of commutativity, distributivity, and so on, but are replaced by different functions. The tropical multiplicative operation is conventional addition, and the tropical additive operation is either min or max; the choice of additive operation distinguishes two isomorphic semirings.  Depending on the choice of min or max as the additive operation, the semiring is given by $\mathbb{T}=(\R\:\cup\{\infty\},\oplus,\otimes)$ or $\mathbb{T}=(\R\:\cup\{-\infty\},\oplus',\otimes)$; $\pm\infty$ are included to serve as additive identities.\footnote{By contrast, the ring of real arithmetic is $(\R,+,\times)$.} These symbols, and others used in this paper, are collected for reference in Table~\ref{t1}. That some of the generating operations of tropical algebra correspond directly to the primitive operations of race logic suggests that it is an ideal setting for the development of time-coded algorithms.\footnote{While tropical algebra is defined over the real numbers with infinity, a race logic circuit can practically represent only a finite discrete set of signal timings. Race logic and tropical algebra are therefore not isomorphic, \textit{per se}. Note that the same is true of a traditional computer with respect to conventional real arithmetic. However, just as traditional computers can operate over a large enough subring of the reals to produce useful calculations, there exists an embedding of min-based race logic as a subsemiring of tropical algebra. Regardless of whether we work in a subset of natural numbers (clocked race logic) or the reals (analog race logic), the fact that this mapping is well-behaved ensures that tropical algebra is a useful mathematical landscape for understanding race logic operations.}

\begin{table}
\centering
\caption{List of symbols related to race logic and tropical algebra and their meanings}
\begin{tabular}{@{}lll@{}} %m{0.8cm}|m{1.5cm}|m{4.5cm}|m{0.8cm}|m{1.5cm}|m{4.5cm}|}
	\toprule
	 Symbol & Meaning  &  Description \\
	 \midrule
	$\infty$ & infinity & additive identity in tropical algebra; edge that never arrives in race logic\\ 
	$\otimes$ & add & multiplicative operation in tropical algebra; temporal delay in race logic\\
	$\oplus$ & min & additive operation in tropical algebra; first arrival in race logic \\
	$\oplus'$ & max & alternate additive operation in tropical algebra; last arrival in race logic \\
	$\dashv$ & inhibit & ramp function in tropical algebra; signal blocking in race logic\\
	$=$ & equivalence & expressing equality between two statements \\
	$:=$ & storage & storing a signal in memory  \\
	$:\cong$ & normalized storage & storing a signal in memory by first performing a normalizing operation \\
	\bottomrule 
\end{tabular}

\label{t1}
\end{table}

Tropical algebra has found numerous applications in the computing literature particularly in a variety of graph algorithms, such as shortest path finding, graph matching and alignment, and minimal spanning trees. It is used as the basis of GraphBLAS (Graph Basic Linear Algebra Subprograms)~\cite{kepner2016mathematical}. In mathematics, it is being used to explore problems in combinatorial optimization~\cite{allamigeon2014combinatorial}, control theory, machine learning~\cite{pmlr-v80-zhang18i}, symplectic geometry~\cite{allamigeon2015tropicalizing}, and computational biology~\cite{yoshida2019tropical}. 
%The main difference between tropical and linear algebra is that the additive operation ($\oplus$) is replaced by either a \textsc{MIN} or a MAX(depending upon the type of algebra) while the multiplicative operation($\otimes$) is replaced by conventional summation. This points to a relationship between such an algebra and race logic, similarities between which we will explore in more detail in the next section. 

%We have specified $\mathbb{T}$ axiomatically, motivated by the obvious connections between race logic and the generating operations of the tropical semiring. Traditionally, $\mathbb{T}$ arises via a limiting process on the familiar ring $(\R\,+,\times)$ of real arithmetic. The basic idea is to represent traditional numbers in an exponential form $t^{x}$ and examine the limit $t\rightarrow 0$; the behavior of the exponents is then governed by the tropical semiring~\cite{}. That tropical algebra can be seen as operating over a logarithmic compression of traditional data may be useful in understanding the comparative usefulness of race logic across different real world datasets.

%An important consequence of tropicalization is the change in the nature of the additive operation. Contrary 
There are some fundamental similarities between tropical algebra and race logic. Both of them have an $\infty$ element. In race logic, it physically corresponds to a signal that never arrives, while in tropical algebra it corresponds to the additive identity, since
\begin{equation}
    \alpha\oplus\infty = \min(\alpha, \infty) = \alpha.
\end{equation}
Such an addition doesn't have an inverse, since there is no value of $\beta$ in $\min(\alpha,\beta)$ that would give $\infty$. The non-invertibility of addition means that this algebra is a semi-ring and fundamentally winner-take-all in nature. Every time the additive operation is performed, the smallest number (the first arriving signal in race logic) ``wins'' and propagates further through the computation.\footnote{If we had chosen $\oplus'$ instead of $\oplus$, the additive identity would be $-\infty$, though we generally prefer the min-plus version of tropical algebra. In pure race logic, $\infty$ corresponds to an edge that never arrived, whereas $-\infty$ would correspond to an edge that had always been present---not to be confused with an edge that arrived at $t=0$. No nontrivial function in pure race logic can output $-\infty$ due to the causality constraint, so the min-plus algebra has considerably more practical utility in race logic.} The multiplicative identity in tropical algebra is zero, rather than one, since $\alpha\otimes0 = \alpha+0=\alpha$.

\

\subsection{Graph problems in tropical algebra}
\label{subsec:graph_trop}

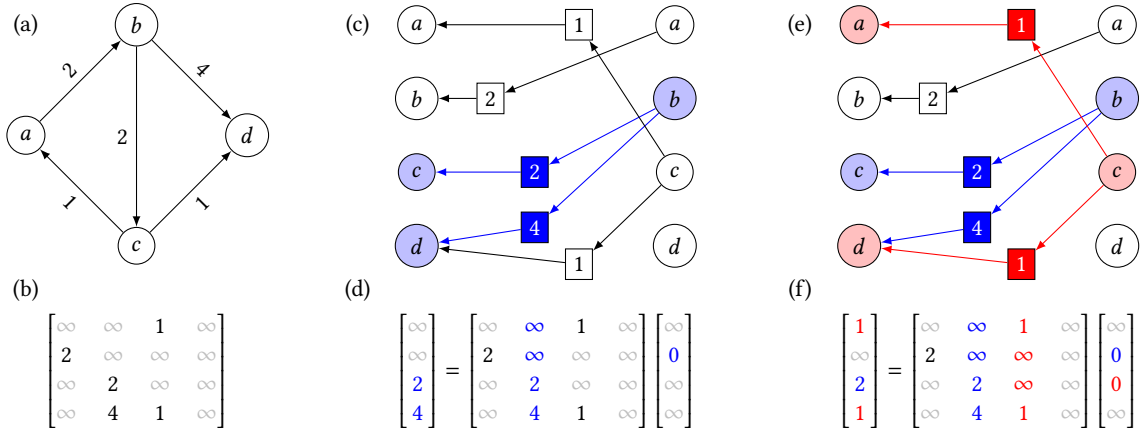
\begin{figure*}
    \centering
\begin{scaletikzpicturetowidth}{\textwidth}
\begin{tikzpicture}[scale=\tikzscale]

\node[] (subfigA) at (-1.5,1) {(a)};
\node[draw,circle] (A) at (-1.5,-0.5) {$a$};
\node[draw,circle] (B) at (0,1) {$b$};
\node[draw,circle] (D) at (1.5,-0.5) {$d$};
\node[draw,circle] (C) at (0,-2) {$c$};

\draw [->,>=latex] (A) -- (B) node [midway,above,sloped] {$2$};
\draw [->,>=latex] (B) -- (D) node [midway,above,sloped] {$4$};
\draw [->,>=latex] (B) -- (C) node [midway,left] {$2$};
\draw [->,>=latex] (C) -- (A) node [midway,below,sloped] {$1$};
\draw [->,>=latex] (C) -- (D) node [midway,below,sloped] {$1$};

\node[] (subfigB) at (-1.5,-2.6) {(b)};
\node[] (M) at (0,-3.7) {
%\begin{displaymath}
$\left[\begin{matrix}
\color{lightgray}\infty& \color{lightgray}\infty & 1 & \color{lightgray}\infty \\ 
2 & \color{lightgray}\infty & \color{lightgray}\infty & \color{lightgray}\infty \\
\color{lightgray}\infty & 2 & \color{lightgray}\infty & \color{lightgray}\infty \\
\color{lightgray}\infty & 4 & 1 & \color{lightgray}\infty
\end{matrix}\right]$
%\end{displaymath}
};

\def \coltwo {4}
\def \colthree {10}

\node[] (subfigC) at ({\coltwo-1},1) {(c)};
\node[draw,circle] (Asrc) at (\coltwo+3.3,1) {$a$};
\node[circle,fill=blue!25!white,text=black,draw=black] (Bsrc) at (\coltwo+3.3,0) {$b$};
\node[draw,circle] (Csrc) at (\coltwo+3.3,-1) {$c$};
\node[draw,circle] (Dsrc) at (\coltwo+3.3,-2) {$d$};

\node[draw] (cToA) at ({\coltwo+2},1) {$1$};
\node[draw] (cToD) at ({\coltwo+2},-2.25) {$1$};
\node[draw] (aToB) at ({\coltwo+0.8},0) {$2$};
\node[draw,fill=blue,text=white,draw=black] (bToC) at ({\coltwo+1.4},-1) {$2$};
\node[draw,fill=blue,text=white,draw=black] (bToD) at ({\coltwo+1.4},-1.75) {$4$};

\node[draw,circle] (Adst) at ({\coltwo-0.2},1) {$a$};
\node[draw,circle] (Bdst) at ({\coltwo-0.2},0) {$b$};
\node[circle,fill=blue!25!white,text=black,draw=black] (Cdst) at ({\coltwo-0.2},-1) {$c$};
\node[circle,fill=blue!25!white,text=black,draw=black] (Ddst) at ({\coltwo-0.2},-2) {$d$};

% Edges of first transition graph
\draw[->,>=latex] (Asrc) -- (aToB);
\draw[->,>=latex,blue] (Bsrc) -- (bToC);
\draw[->,>=latex,blue] (Bsrc) -- (bToD);
\draw[->,>=latex] (Csrc) -- (cToA);
\draw[->,>=latex] (Csrc) -- (cToD);
\draw[->,>=latex] (aToB) -- (Bdst);
\draw[->,>=latex,blue] (bToC) -- (Cdst);
\draw[->,>=latex,blue] (bToD) -- (Ddst);
\draw[->,>=latex] (cToA) -- (Adst);
\draw[->,>=latex] (cToD) -- (Ddst);

\node[] (subfigD) at ({\coltwo-1},-2.6) {(d)};
\node[] (M) at ({\coltwo+1.55},-3.7) {%\begin{displaymath}
$\left[\begin{matrix}
\color{lightgray}\infty\\
\color{lightgray}\infty\\
\color{blue} 2\\
\color{blue} 4
\end{matrix}\right]
=\left[\begin{matrix}
\color{lightgray}\infty& \color{blue}\infty & 1 & \color{lightgray}\infty \\ 
2 & \color{blue}\infty & \color{lightgray}\infty & \color{lightgray}\infty \\
\color{lightgray}\infty & \color{blue}2 & \color{lightgray}\infty & \color{lightgray}\infty \\
\color{lightgray}\infty & \color{blue}4 & 1 & \color{lightgray}\infty
\end{matrix}\right]
\left[\begin{matrix}
\color{lightgray}\infty\\
\color{blue}0\\
\color{lightgray}\infty\\
\color{lightgray}\infty
\end{matrix}\right]$};
%\end{displaymath}};

%%% Righthand column (two-hot activation)

\node[] (subfigE) at ({\colthree-1},1) {(e)};
\node[draw,circle] (AsrcR) at (\colthree+3.3,1) {$a$};
\node[circle,fill=blue!25!white,text=black,draw=black] (BsrcR) at (\colthree+3.3,0) {$b$};
\node[circle,fill=red!25!white,text=black,draw=black] (CsrcR) at (\colthree+3.3,-1) {$c$};
\node[draw,circle] (DsrcR) at (\colthree+3.3,-2) {$d$};

\node[fill=red,text=white,draw=black] (cToRA) at ({\colthree+2},1) {$1$};
\node[fill=red,text=white,draw=black] (cToRD) at ({\colthree+2},-2.25) {$1$};
\node[draw] (aToRB) at ({\colthree+0.8},0) {$2$};
\node[fill=blue,text=white,draw=black] (bToRC) at ({\colthree+1.4},-1) {$2$};
\node[fill=blue,text=white,draw=black] (bToRD) at ({\colthree+1.4},-1.75) {$4$};

\node[circle,fill=red!25!white,text=black,draw=black] (AdstR) at ({\colthree-0.2},1) {$a$};
\node[draw,circle] (BdstR) at ({\colthree-0.2},0) {$b$};
\node[circle,fill=blue!25!white,text=black,draw=black] (CdstR) at ({\colthree-0.2},-1) {$c$};
\node[circle,fill=red!25!white,text=black,draw=black] (DdstR) at ({\colthree-0.2},-2) {$d$};

% Edges of first transition graph
\draw[->,>=latex] (AsrcR) -- (aToRB);
\draw[->,>=latex,blue] (BsrcR) -- (bToRC);
\draw[->,>=latex,blue] (BsrcR) -- (bToRD);
\draw[->,>=latex,red] (CsrcR) -- (cToRA);
\draw[->,>=latex,red] (CsrcR) -- (cToRD);
\draw[->,>=latex] (aToRB) -- (BdstR);
\draw[->,>=latex,blue] (bToRC) -- (CdstR);
\draw[->,>=latex,blue] (bToRD) -- (DdstR);
\draw[->,>=latex,red] (cToRA) -- (AdstR);
\draw[->,>=latex,red] (cToRD) -- (DdstR);

\node[] (subfigF) at ({\colthree-1},-2.6) {(f)};
\node[] (M) at ({\colthree+1.55},-3.7) {
%\begin{displaymath}
$\left[\begin{matrix}
\color{red}1\\
\color{lightgray}\infty\\
\color{blue} 2\\
\color{red} 1
\end{matrix}\right]
=\left[\begin{matrix}
\color{lightgray}\infty& \color{blue}\infty & \color{red}1 & \color{lightgray}\infty \\ 
2 & \color{blue}\infty & \color{red}\infty & \color{lightgray}\infty \\
\color{lightgray}\infty & \color{blue}2 & \color{red}\infty & \color{lightgray}\infty \\
\color{lightgray}\infty & \color{blue}4 & \color{red}1 & \color{lightgray}\infty
\end{matrix}\right]
\left[\begin{matrix}
\color{lightgray}\infty\\
\color{blue}0\\
\color{red}0\\
\color{lightgray}\infty
\end{matrix}\right]$};
%\end{displaymath}};
\end{tikzpicture}
\end{scaletikzpicturetowidth}
    \caption{Tropical matrices for graph exploration: (a) shows an example directed graph; (b) shows the equivalent weighted adjacency matrix. Panel (c) shows the propagation of a signal originating at node $b$ through a delay network corresponding to the edges of the example graph. Panel (d) shows the tropical vector-matrix multiplication corresponding to panel (c). Panels (e) and (f) repeat these representations for the case where signals are injected at both $b$ and $c$.}
    \label{fig:2graph_trav}
\end{figure*}

%Tropical algebra can be especially useful for graph analytics, where it provides a simple mathematical language for graph traversal. It is used for a range of graph problems, particularly the single-source shortest path problem and its derivatives such as all-pair shortest paths, closeness centrality, and betweenness centrality. In this section we describe the mapping between a tropical vector-matrix multiplication and a graph traversal operation, giving an intuitive interpretation of tropical vector algebra. 
Tropical algebra can be especially useful for graph analytics, where it provides a simple mathematical language for graph traversal. A fundamental concept in graph traversal is the graph's weighted adjacency matrix $A$. Figs.~\ref{fig:2graph_trav}(a) and (b) show a directed graph and its weighted adjacency matrix, respectively. The $i^\text{th}$ column of the weighted adjacency matrix represents the distances of the outward connections from node $i$ to all other nodes in the graph, so that $A_{ji}$ is the weight for the edge $i\rightarrow j$. Where there is no edge to node $i$ from $j$, we assign the value $A_{ji}=\infty$. %This representation of a graph is quite standard and has not yet invoked any tropical interpretation.

%\begin{figure*}[h]
%    \centering
%    \includegraphics[width=\linewidth]{Figures/2_graph_trav.pdf}
%    \caption{Tropical Matrices for Graph exploration: (a) shows an example directed graph while panel (b) shows the equivalent weighted adjacency matrix. Panel (c) shows the operation of a left row multiplication of a tropical one-hot distance vector with the weighted adjacency matrix resulting in the the tropical output vector shown below it. Panel (d) shows the same with a different input vector. The colored rows in panel (c) and (d) represent the interaction of the elements in the row vector with the rows in the adjacency matrix during a tropical vector-matrix multiple (VMM) operation. Panels (e) and (f) show an alternate representation of the graph  where the circles represent the nodes while the boxes on the edges represent the weights which can be seen as delay elements. This interprets the behaviour of tropical matrix multiplication as temporal exploration.   }
%    \label{fig:2graph_trav}
%\end{figure*}

The usefulness of tropical algebra for graph traversal is seen when using $A$ in a tropical vector-matrix multiplication. Tropical vector-matrix multiplication (VMM) proceeds like conventional VMM, but with $(\oplus,\otimes)$ instead of $(+,\times)$. As shown in Fig.~\ref{fig:2graph_trav}, each vector element is scaled (tropical multiplication) before they are all accumulated (tropical addition). Extracting any single column from a matrix can be done by multiplying a one-hot vector as shown in Fig.~\ref{fig:2graph_trav}. The tropical one-hot vector has a single zero element with all other entries set to $\infty$; from Sec.~\ref{subsec:trop_bckgnd}. During scaling, the columns of the adjacency matrix that correspond to the infinities of the one-hot vector get scaled to infinity (tropically multiplied by $\infty$) while the remaining column, scaled by the multiplicative identity $0$, is the output. The values stored in the output vector represent the distances from the one hot node in the input vector. This operation represents a search from the node in question (decided by the one-hot vector) to all the connected nodes in the graph, and reports the distances along all edges of this parallel search.

Using a ``two-hot'' vector for input, as shown in Fig.~\ref{fig:2graph_trav}(d) outputs a tropical linear combination of two vectors, corresponding to the ``hot'' columns of the adjacency matrix. The accumulation phase of the tropical VMM is nontrivial; the $\oplus$ operation selects the smallest computed distance to each node for the output. The tropical VMM reports the shortest distance to each node in the graph after a single edge traversal from \emph{either} of the initial nodes specified by the two-hot vector. Both steps---the exploration of a node's neighbors and the elementwise minimum of possible parent nodes associated with an output---are performed in parallel by a single matrix operation. %Each tropical VMM operation is equivalent to performing a single parse along all edges from all nodes while calculating the shortest distances at the end of the parse -- a pretty powerful operation.

Representing a collective hop through the graph as a single matrix operation allows a series of matrix operations to represent extended graph traversal. The shortest traversed distance to each node in a graph from an initial node $x$ is  %This allows dynamic programming to be expressed quite easily as follows 
\begin{equation}
    y = x\oplus (x \otimes A) \oplus (x \otimes A \otimes A) \oplus (x \otimes A \otimes A \otimes A) \oplus \cdots .
\end{equation}
The first term represents all single-hop shortest paths starting out from $x$, while the second term accounts for all the two-hop shortest paths, and so on. Hence the tropical summation across all the terms in $y$ allows it to encode the shortest distances between the input node as specified by $x$, independent of the number of hops. Performing $N$ such hops and calculating the minimum distance across all of them is the key operation in various dynamic-programming-based shortest path algorithms. This process makes tropical algebra the natural semiring for Dijkstra's algorithm~\cite{mohri2002semiring}. We use these ideas to implement Dijkstra's single-source shortest path algorithm in a stateful race logic system in Sec.~\ref{sec:dijkstra}.

\subsection{Circuits for tropical linear algebra}
\label{subsec:rl_trop}

 %Tropical linear algebra can thereby serve as an energy efficient foundation for higher order race computing, and our goal now is to understand how to build the circuit primitives needed for such a system.

Since tropically linear functions $\bigoplus_j (a_j\otimes t_j)$ with the $a_j$ values constant satisfy the invariance condition, tropical linear transformations may be carried out in pure race logic. In Section~\ref{subsec:rl_bckgnd}, we describe how single rising edges can be used to encode information in their arrival time. Interpreting the edges as tropical scalars, we can see how \textsc{or} gates and delay elements are modeled by tropical addition and multiplication. This understanding can also be extended to tropical vectors. Section~\ref{subsec:graph_trop} describes how tropical vectors can be interpreted as distance vectors in graph operations. These distance vectors can be interpreted temporally as a wavefront or a volley of edges measured with respect to a temporal origin. Other researchers have proposed using such vectors as the primary data structure underlying temporal computations \cite{smith2018space,tavanaei2018deep}.

%Tropical vectors and temporal wavefronts share other interesting properties, which arise by virtue of tropical multiplication being similar to conventional addition. 
Just as conventional vectors can be normalized, so can tropical vectors. In tropical algebra, the vector norm is the minimum element of the vector.\footnote{In the max-plus tropical semiring, the vector norm would be the maximum element of the vector. This vector magnitude operation is sometimes called the $L^\infty$ norm.} Tropical division of a vector by its norm then corresponds to subtracting out this minimum value from all the components. This ensures that at least one element of the tropical vector is zero. It is common to regard a tropical vector as equivalent to its normalized version, implying that it only encodes information about the shape of its temporal wavefront, and not about an arbitrarily chosen temporal origin. To accept this equivalence is to work in the \emph{projective} tropical vector space, and we refer to tropical vectors as being projectively equivalent if their normalized versions are elementwise equal. In this paper we frequently make use of normalized tropical vectors and describe a method to store vectors projectively.  Not only are they commonly the naturally correct data structure, but frequent renormalization of the temporal origin helps mitigate the limited dynamic range of our  temporal encoding. Keeping track of the relative normalization constants allows us to encode information that would nominally extend beyond our dynamic range in a principled way.

\begin{figure*}[]
    \centering
    \includegraphics[width=\linewidth]{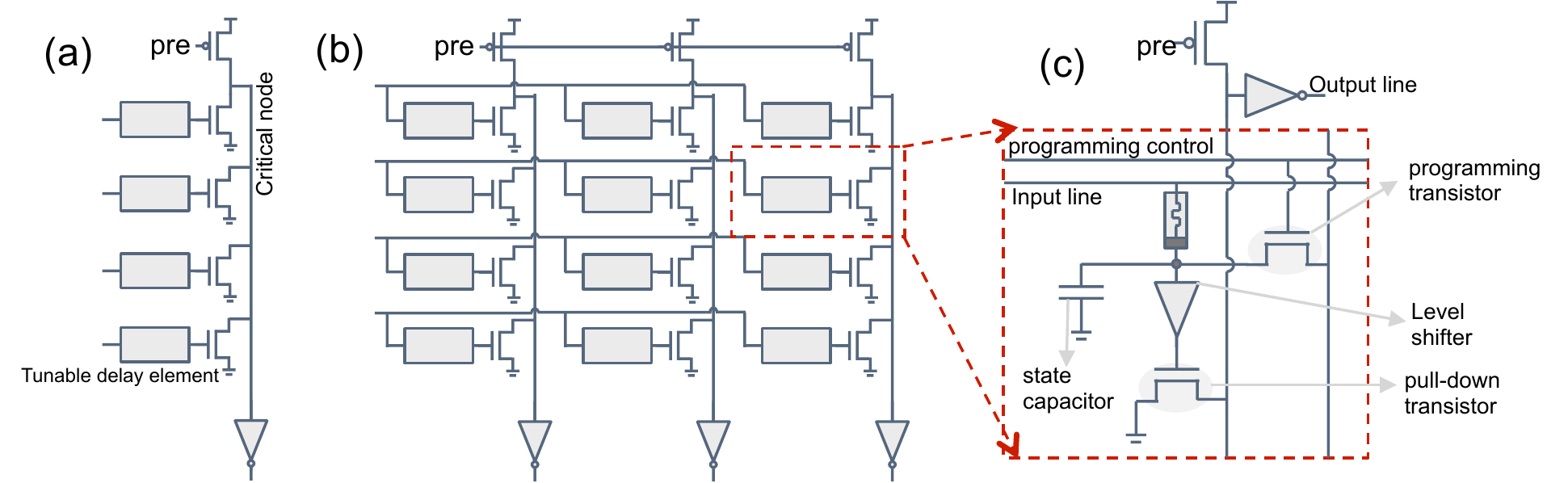}
    \caption{Construction of race logic circuits for tropical algebra: Panel (a) shows a composite circuit for the tropical dot product operation.   A simple array of delay elements  takes an incoming wavefront and delays its elements by the values stored in each of the delay elements. This represents the tropical element-wise multiplication by constant operation. The output is then connected to a $p$-type, metal-oxide semiconductor (PMOS) pre-charge pullup coupled with a \textsc{nor}-style pulldown network which behaves like a first arrival detector and performs the tropical addition operation. Panel (b) combines multiple elements of panel (a) and scales this up to a 2D array such that it performs tropical vector matrix multiplication, the critical operation for graph traversal as described in section \ref{subsec:graph_trop}. Panel (c) shows a detailed circuit implementation of the tropical VMM cell. Each cell consists of two transistors, one for programming and the other for operation, and a level shifter. In the programming mode, the array is used like a conventional 1T1R array and the memristors are written to the appropriate resistance values. In the operation mode, the programming transistor is turned off, while the gate capacitor (shown in figure) is charged through the memristor. The level shifter is used to make sure that the discharge time constant is determined by the memristor charging process and not the pulldown of the transistor, by applying full swing inputs to the pulldown transistor.}
    \label{fig:3tropops}
\end{figure*}

Once a wavefront of rising voltage edges is interpreted as a tropical vector, the techniques shown in Fig.~\ref{fig:3tropops} can be used to implement tropical vector operations. Panel (a) shows the vectorized version of the tropical dot product operation. First the column of delay elements delays each line in the incoming wavefront by a different amount. This implements tropical multiplication by constants, and can be seen as superimposing the delay wavefront onto the incoming wavefront. The outputs of such a circuit are then connected to the inputs of a pre-charge-based pullup with an \textsc{or}-type pulldown network followed by an inverter. The circuit operation is divided in to two phases, the pre-charge phase followed by the evaluation phase. In the pre-charge phase, the PMOS transistor has its input connected to ground, causing the critical node to be pulled-up~(connected to $V_{dd}$). When the pre-charge phase ends, the PMOS transistor is turned off, which maintains the potential at the critical node at $V_{dd}$. During the evaluation phase, the the first arriving rising edge at the input of the one of the NMOS transistors, causes the critical node to discharge to ground, hence being pulled-down to a potential of zero volts. This behaves as a first-arrival detection circuit that outputs a rising edge at the minimum of the input arrival times, performing the $\min$ operation. It implements tropical vector addition. Combining the delay (multiplication) with the min (summation), we get the tropical dot product operation. By replicating this behavior across multiple stored vectors, as in panel (b), we get the tropical VMM operation, where the input vector tropically multiplies a matrix.

To be specific, we consider versions of the tunable delay elements described in the previous paragraph that are based on memristor or resistive random access memory (ReRAM) technology. In the tropical VMM such a device is used as programmable resistor with a known capacitance to generate an $RC$ delay~\cite{madhavan2020storing}. The details of these tropical vector algebra cells are shown in Fig.~\ref{fig:3tropops}(c).  The main element of this circuits is a 2T1R array comprised of a pulldown transistor and a programming transistor.  During the programming phase, the programming transistor coupled with the programming lines can be used to apply the necessary read and write voltages across the memristor, thus changing the resistance and therefore $RC$ delay time stored in the device. During the operation of the circuit, the programming transistor is turned off to decouple the programming lines from the active circuitry. In the pre-charge phase, the output lines are pulled up to $V_\text{dd}$ through the pullup transistor. In the evaluation phase, the input lines receive temporally coded rising edges which charge up the gate capacitors as shown in Fig.~\ref{fig:3tropops}. This causes the pulldown transistor to be turned on at the time proportional to input arrival times plus the $RC$ time constant of the coupled memristor-capacitor in each cell, faithfully performing the tropical VMM operation. 

The largest read voltage that can applied across the device without disturbing the state of the device is approximately 600~mV. In a 180~nm process, this value is only a few 100~mV above the transistor threshold voltage and would cause a slow and delayed leak. This leak allows multiple inputs to affect the pulldown simultaneously, influencing the functional correctness of the circuit. We propose two solutions to this problem. Figure~\ref{fig:3tropops}(c) shows a level shifter added between the memristor and the pulldown transistor, the full swing of which causes the pulldown transistor to work much faster. In an alternate approach (not shown here), a medium $V_{th}$ device is used for the pulldown. Such devices ensure a small footprint as well as correct operation, provided the fabrication process allows them. 

In addition to tropical VMM based linear transformations, other primitive vector operations are crucial in many real applications. Elementwise min and max can be performed with arrays of \textsc{or} and \textsc{and} gates, respectively. Vectors can also be reduced to scalars by computing min or max amongst all elements using multi-input \textsc{or} and \textsc{and} gates. 

% \begin{figure}
%     \centering
%     \includegraphics[width=\linewidth]{Figures/4_temporal_tropical_vmm.pdf}
%     \caption{Temporal tropical-VMM circuits with memristors: Panel (a) shows the general layout as shown in figure \ref{fig:3tropops}. Panels (b) and (c) show two different detailed circuit implementations of the tropical VMM. In both panels, each cell consists of two transistors, one for programming and the other for operation. In the programming mode, the array is used like a conventional 1T1R array and the memristors are written to the appropriate resistance values. In the operation mode, the programming transistor is turned off, while the gate capacitor (shown in figure) is charged through the memristor. In panel (b) a level shifter is used to make sure that the discharge time constant is determined by the memristor charging process and not the pulldown of the transistor, by applying full swing inputs to the pulldown transistor. Panel (c) shows an alternate approach using a low $V_{th}$ transistor.}
%     \label{fig:4tropvmm}
% \end{figure}

%\subsection{Tropical VMM Circuit}
%\begin{itemize}
%    \item How does it work? with circuit details.
%    \item multiple modes of operation: vectoring mode and co-vectoring mode. (produce a vector from an input scalar, or produce a scalar (dot product) from an input vector)
%    \item How it can be used to perform tropical VMM operations. 
%\end{itemize}

\subsection{Circuits for nonlinear tropical functions}
\label{subsec:rl_other}

\begin{figure}
    \centering
    \includegraphics[width=\linewidth]{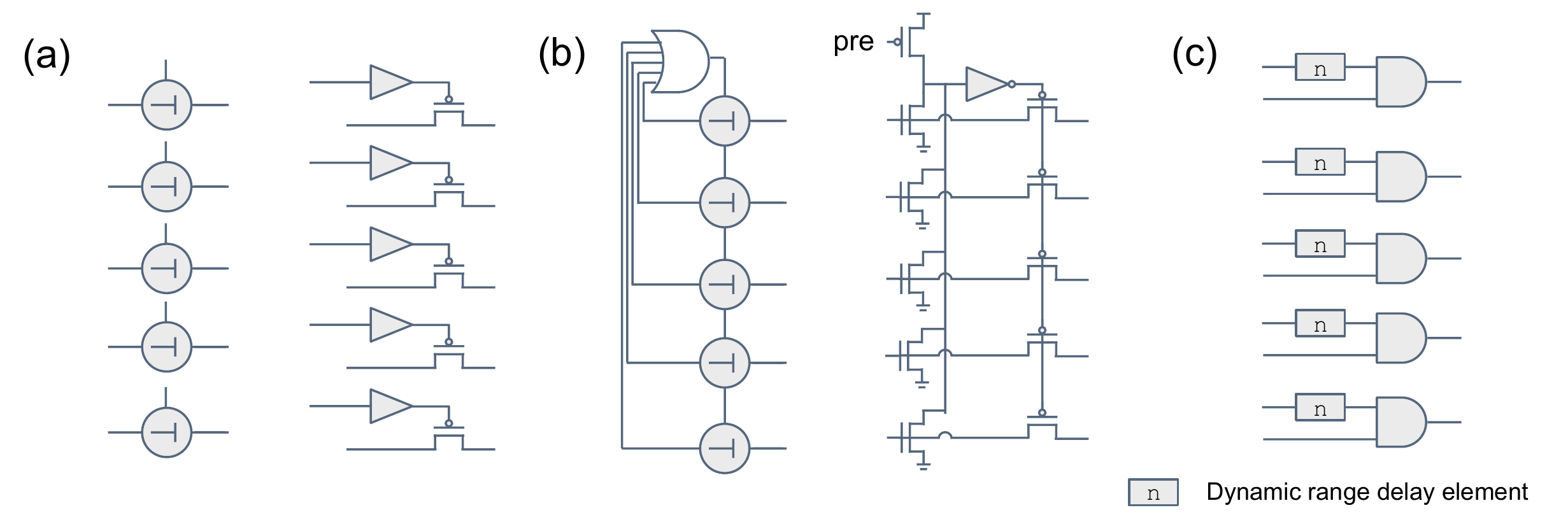}
    \caption{Tropically nonlinear race logic functions: Panel (a) shows the conceptual and circuit diagram for an element-wise inhibit operator. The inhibiting input is buffered before being fed into the gate terminal of a PMOS. As the inhibiting input turns high, the PMOS turns off inhibiting the secondary input. Panel(b) shows the argmin operation that takes an input vector and returns a one-hot vector at the location of the element with the minimum value. This is done by taking the first arrival signal and inhibting everything else but that signal. Panel (c) shows a binarizer. An input wavefront is maxed with the all $n$ wavefront. This takes all values to this max value, except $\infty$, which remains as is, performing binarization.}
    \label{fig:5transfunc}
\end{figure}

Apart from circuits that allow race logic to implement tropical linear algebra, additional built-in functions, such as elementwise inhibit, argmin, and binarization, are required to perform general purpose tropical computations. Elementwise inhibit, shown in Fig.~\ref{fig:5transfunc}(a), is particularly powerful, as it allows us to implement piecewise functions. Its technical operation follows directly from the scalar inhibit operation discussed in Sec.~\ref{subsec:rl_bckgnd}.

The argmin function, shown in Fig.~\ref{fig:5transfunc}(b), converts its vector input to a tropical one-hot vector that labels a minimal input component. An \textsc{or} gate is used to select a first arriving signal which then inhibits every vector component. Only one first arriving edge survives its self-inhibition; no other signals in the wavefront are allowed to pass, effectively sending these other values to infinity. The resulting vector is projectively equivalent to a tropical one-hot, and achieves the canonical form with a single zero among infinities after normalization.\footnote{A variety of conventions could be taken for the case where more than one signal arrives at the same, earliest time. Note that such a multi-hot vector can be generated by the circuit shown in Fig.~\ref{fig:5transfunc}(b). When such a situation occurs, a sorted delay vector can be used to select one of the hot input elements and convert the vector to a one-hot vector.}

The binarization operation, shown in Fig.~\ref{fig:5transfunc}(c), is similar; it converts all finite components to $0$ while preserving infinite components at $\infty$. This operation utilizes a pre-stored vector which has the maximum finite (non-$\infty$) value $t_\text{max}$ of the computational dynamic range on each component. We define $\text{binarize}(\vec x) = t_\text{max} \oplus' \vec x$. Computing the elementwise max of such a vector with any incoming vector, values that are $\infty$ remain so while the other values are converted to the maximal finite input value. Normalizing the result via projective storage saves a many-hot vector labeling the finite components of the original input.

\section{Temporal state machines}
\label{sec:temporal-state-machine}
%\begin{itemize}
%    \item In order to solve problems of various sizes and types, we need to be able to effectively partition any problem into smaller sub-problems which can be stitched together effectively. Here will need notions of state and state transition functions. [make sure to include "state" machines as we have discussed them.]

%    \item Take an example of a simple data flow computation with one add, one min and one inhibit. Show what that looks like and how it can be implemented with our proposed state machine. 
%\end{itemize}

The \textit{finite state machine} or \textit{finite state automaton} is a central concept in computing and lies at the heart of most modern computing systems. Such a machine is in one of some finite set of states $S$ at any particular instant in time; inputs $x\in\Sigma$ to the machine both produce outputs $y\in\Gamma$ and induce transitions between these internal states. A \emph{state transition function} $\delta : S\times\Sigma \rightarrow S$ determines the next state based on the current state and current input, and an \emph{output function} $\omega : S\times\Sigma \rightarrow \Gamma$ gives the output based on the state and inputs of the machine.\footnote{This specification of a state machine is called a Mealy machine; if $\omega$ depends only on the state and not the current input, it is called a Moore machine. The two models are equally powerful in principle.}

The presence of state means that there is not a one-to-one correspondence between input and output of the machine; in the language we have developed above, a state machine is an impure function. This impurity is due entirely to the state variable; $\delta$ and $\omega$ are pure mathematical functions. The finite state machine thus provides a template for how we might compose pure race logic functions together across stateful interfaces to create general purpose temporal automata. In fact, the temporal state machines we introduce below fits into the mathematical framework given above.

%It includes two major components, namely memory and logic. State machines have memory units that hold state and represent current state of the computation; this memory is usually implemented with flip flops or registers. They also have combinational logic units that provide the state transition and output functions; this logic is conventionally implemented with Boolean logic gates. 

The temporal state machine we introduce differs from conventional automata in that the signals use temporal rather than Boolean encoding. State is made possible by recent proposals for temporal memories. These memories use temporal wavefronts as their primary data structure. By coupling nanodevice parameters to pulse duration, they are able to freeze temporal data in device properties such as resistance. Together with the pure race logic primitives described in previous sections, we can now build finite state automata in an end-to-end temporal encoding. 

Designing such a machine requires addressing several problems intrinsic to the temporal nature of these logic and memory primitives. We start this section with a brief background on temporal memories, based on hybrid CMOS and emerging technologies, and explain their benefits and drawbacks. Then we describe the impure tropical multiplication of two signals in race logic as a first example of composing pure race logic across stateful interfaces in order to break through the invariance restriction. Finally, we return to the general state machine formulation and argue for the extensibility of our simple example to more complex systems.

\subsection{Temporal Memory}
\label{subsec:tm_bckground}
\begin{figure}
     \centering
     \includegraphics[width=\linewidth]{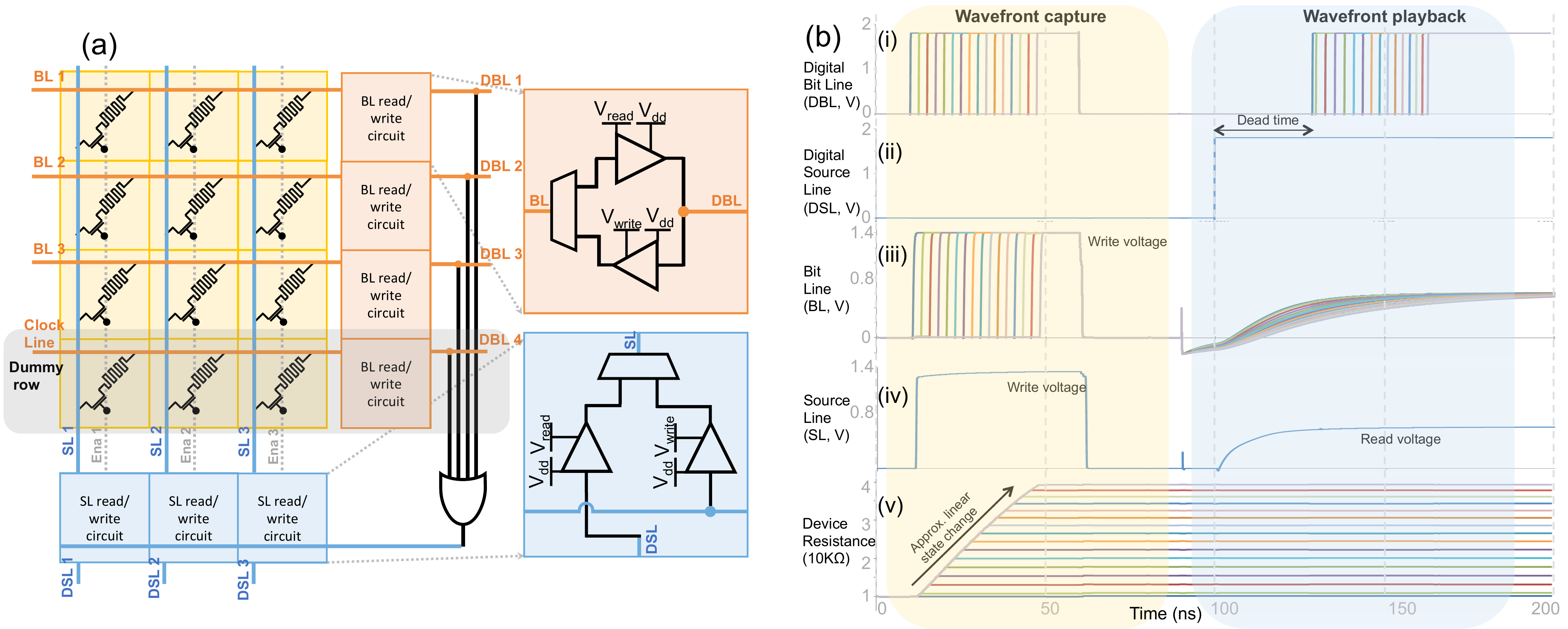}
     \caption{Memristive wavefront memory: Panel (a) shows a $4\times 4$ memristive temporal memory, complete with read and write peripheral circuits as described in Ref.~\cite{madhavan2020storing}. Note that bit-line-4 has been replaced by a dummy line where the resistance values are fixed. The time constant of this line is governed by the parasitics of the circuit and determines the temporal origin of the outgoing wavefront. . Panel (b) shows the functioning of a $16\times 16$, 4-bit, temporal memory as simulated in our 180~nm Silterra process. Strip (i) shows the capture and playback of a linearly varying digital wavefront, with each color representing one of the sixteen lines involved. These edges have been collapsed into a single strip for clarity.  Note that small timing mismatches cause small changes in the shape of the wavefront that is played back. Strip (ii) shows in the digital read input applied to a captured column. Strips (iii, iv) show the source lines and bit lines, but internal to the memory and hence operate at different voltages which are shifted to $V_{dd}$ with level shifters as shown in panel(a). Strip (v) shows the state change behaviour of the memristors as given by the memristor model in \cite{chen2015compact}. Note that the state change is almost linear. Careful inspection reveals a slight convexity by virtue of higher order terms in the exponential dependence.}
     \label{fig:6tempmem}
\end{figure}

Temporal memories natively operate in the time domain. They operate on wavefronts of rising edges rather than on Boolean values. Such memories can be implemented with emerging technologies such as memristors (as shown in~Fig.\ref{fig:6tempmem}) and magnetic race tracks~\cite{vakili2020temporal} because the physics includes a direct coupling between the time variable and some analog device property. In any case, the memory structures are similar: memory cells are arranged in a crossbar. For the read operation, a single rising edge represented by the tropical one-hot vector is applied to the input address line of the memory, creating a wavefront at the output data line. For a write operation, the column of the crossbar where the memory has to be stored is activated and an incoming wavefront is captured. 

The way temporal information is encoded in the devices depends on the technology. For memristors, the dependence of the $RC$ charging time constants on the resistance $R$ of the memristor and the row capacitance $C$ is used to encode the temporal information, leading to a linear relationship between timing and resistance. Utilizing a 1T1R-like structure, the shared row capacitances are the output capacitances that have to be charged. In the write operation the temporal dependence of state change of the device creates a relative change in resistive state based on arrival time of the edges. This enables a set of devices to correctly encode the shape of an incoming wavefront. 

An alternative proposal in the literature uses magnetic racetracks to store temporal information in the position of a magnetic domain within each track~\cite{vakili2020temporal}. Magnetic tracks have the property that a current passing through a metal layer below the magnet causes motion of the magnetic domain in the direction of the current flow. The speed $v$ of magnetic domain motion is constant for constant current amplitude, and so the simple equation $x=vt$ determines the final location of the domain. This linear proportionality provides a straightforward mapping between the timing of a write signal and the position of the stored magnetic domain. %This relationship is easily exploited for both the read and the write operations.  

Regardless of the technology, the behaviour of these memories is qualitatively different from that of registers or flip-flops. In the case of registers, a single clock tick performs two functions. Not only does it capture the next state information from the calculation performed in the previous cycle, it also initiates the next cycle's computation. Combinational logic is thus ``stitched'' together by register interfaces. This feature of conventional memory does not exist in the temporal memories proposed to date because wavefront playback and capture use the same address and data lines, and cannot be used at the same time. Addressing this deficiency requires memories that can be used both upstream and downstream for the same operation, as shown in Fig.~\ref{fig:6tempmem}.

% \begin{figure}
%     \centering
%     \includegraphics[width=\linewidth]{Figures/7_addition_state_machine.pdf}
%     \caption{Tropical multiplication with a temporal state machine: Panel (a) shows the structure of the temporal state machine, with simple control signals not shown for clarity. It consists of three memory units which can be configured for reading or writing, and a temporal arithmetic unit which is composed of invariant race logic primitives combined with a single column temporal memory for addition. Panel (b) shows the additive unit inside the temporal arithmetic unit, as well as two different phases that are involved in the temporal addition(tropical element-wise multiplication process). The first phase of operation is to store an incoming wavefront in a temporal memory through the write circuitry. In the next phase the stored wavefront is played over the input wavefront, hence correctly performing temporal addition.  }
%     \label{fig:7_add_state_mach}
% \end{figure}

Some limitations of the temporal memories discussed above arise because they are analog\footnote{The computing scheme discussed here can be either analog or digital. Though our evaluation (Sec.~\ref{sec:performance-results}) is done assuming analog behavior, noise and other non-idealities will in practice determine the information capacity afforded to such a computing scheme. We discuss this issue further in Sec.~\ref{sec:comp_tech_considerations}. }: they possess limited dynamic range, and a dead time is incurred in their use as shown in~\ref{fig:6tempmem}(b). The dead time is as a result of the charging of the parasitics of the array, which---with growing array size---can become comparable to the delays stored. As measured from the temporal origin of the calculation, the dead times introduce artificial delays in each component that result in incorrectly encoded values at a memory write input. To deal with this issue, we introduce an extra dummy line, which we call the \emph{clock line}, that always has the minimum $R_\text{on}$ value for the resistor. This line serves as a temporal reference to the origin and hence behaves like a clock. This ensures that the parasitics of the lines are accounted for and only the relative changes in the resistance values are translated to the output wavefront. 

The dynamic range is determined by the relative changes in the stored resistances, which manifest themselves as changes in the shape of the wavefront with respect to the clock line. Even optimistically, the range is limited to 6 bits to 7 bits with present technologies. Given our constrained dynamic range, we often restrict ourselves to the storage of normalized tropical vectors. In Sec.~\ref{subsec:trop_bckgnd}, we describe how this can be achieved by subtracting from each component the minimal value among all components. This guarantees that at least one element is zero in the normalized result. This alters the reference time of the calculation by the normalization constant which was subtracted away. Some algorithms are insensitive to this shift; otherwise, the normalization constant can be stored in an additional memory element for later recovery. In order to store the normalized version of a tropical vector, the min value of the vector (without the clock line) is used to replace the clock line for the storage operation, re-assigning it as the temporal origin. This can be performed by pulling the clock line input in Fig.~\ref{fig:6tempmem}(a) to $V_{dd}$.

\subsection{Invariance and temporal addition}
\label{subsec:tm_addition}

In Section~\ref{subsec:rl_bckgnd}, we describe the invariance restriction on pure race logic. It constrains stateless race logic circuits to the computation of tropically linear functions. An immediate consequence is that pure race logic cannot tropically multiply two temporal signals. Static delay elements can be used to increment the value of a temporal signal by some fixed amount, but the raw addition of two time codes $t_1+t_2$ is physically forbidden in the presence of time-translational symmetry.\footnote{Under the invariance constraint (Sec.~\ref{subsec:rl_bckgnd}), if two signals $t_A$ and $t_B$ are both shifted by a constant time $\delta$, the output of a function of those signals must also be shifted by the same amount, so that $f(t_A+\delta,t_B+\delta) = \delta + f(t_A,t_B)$. This doesn't work for addition: if $f(t_A,t_B)=t_A+t_B$, shifting the inputs would result in $t_A+t_B+2\delta$ at the output. Therefore addition is not temporally invariant and two temporal signals cannot be added together using pure race logic.} We can break this symmetry in stateful race logic through the introduction of memory.%, the writing to which naturally introduces privileged spacetime events.
%When race logic is restricted to only temporally invariant operations it cannot implement all processes described by tropical algebra basically because two temporal signals cannot be added together cite{jim,truth-matrix}.  This property, in a temporal encoding scheme restricts the kinds of computation that can be performed. This is the reason why one of the four fundamental primitives is an ADD-BY-CONSTANT and not the addition of two signals, as the latter would break temporal invariance. Specifically, if two signals $A$ and $B$ are both shifted by a constant time $c$, the output should also be shifted by the same amount. As one can see, this doesn't work for addition as a faithful addition would result in $A+B+2c$ as the output, which would not be temporally invariant. Not being allowed to add signals is a major impediment. Most importantly, it causes applications where addition is necessary, to be arranged in a feed-forward chain of computational elements that have to be pre-programmed and selected through multiplexing.

With temporal memory, tropical multiplication of two wavefronts proceeds by breaking the operation into two phases as shown in Fig.~\ref{fig:8sm_oper}(a,b). The first panel shows the first phase which stores the incoming wavefront in a local temporal memory using wavefront capture circuits. This stored vector can be temporally added to a new incoming wavefront, as shown in the next panel. Commutativity ensures that the order of storage and playback does not matter. Though the state transition and output functions within each phase are pure race logic functions, the state breaks invariance across the phase boundaries. Using memory for tropical multiplication thus allows us to construct tropical multinomial functions of arbitrary order.

\subsection{A sample temporal state machine} 
\label{subsec:tm_toy}

\begin{figure*}
    \centering
    \includegraphics[width=\linewidth]{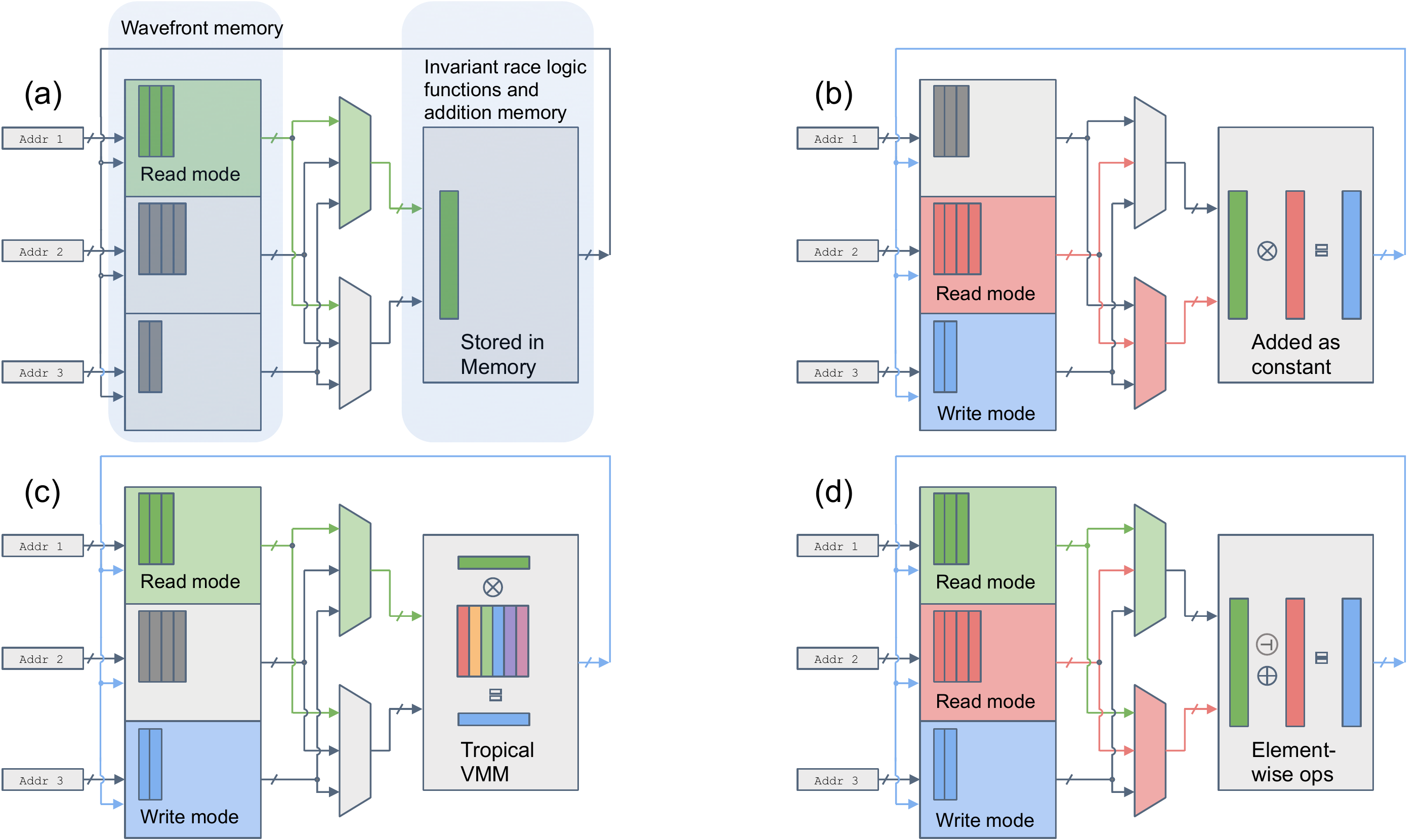}
    \caption{State machine operations: The state machine is partitioned into two main units: the temporal wavefront memory and the arithmetic unit. These are shown in panel (a). The multiplexers and the read/write modes of the memory allow the operations to be performed sequentially. Depending on the operations, individual memory units can behave as either upstream or downstream memories. Panels (a) and (b) show tropical multiplication in a temporal state machine split into two state transitions. In panel (a) storage of the incoming wavefront manifests as a one-argument operation; the vector is stored in the additive memory bank. The next phase in panel (b) is another one-argument operation, where the incoming wavefront is delayed by the wavefront stored in the previous phase. Panel (c) shows the tropical VMM operation. Panel (d) shows other element-wise operations that can be performed in a temporal state machine. Note that all operations, aside from the first phase of the tropical multiplication, store an output back in the temporal memory. The element-wise operations are the only two-argument operations and involve all three memories: the read memories are the upstream memories, while the write memory is the downstream memory.}
    \label{fig:8sm_oper}
\end{figure*}

The invariant race logic circuits and temporal wavefront memory described above are sufficient to build a simple temporal state machine, as shown in Fig.~\ref{fig:8sm_oper}(a). It consists of three banks of temporal memory, which can receive address inputs from external sources as well as data inputs from the output of the machine. The data outputs of the wavefront memory are multiplexed into the computation unit. This unit consists of a variety of the invariant race logic functions from Sec.~\ref{subsec:rl_bckgnd} as well as a temporal memory unit for tropical VMM described in Sec.~\ref{subsec:rl_trop}. The structure allows for a maximum of two-operand operations to be executed at once. 

\begin{figure}
\removelatexerror
\begin{algorithm}[H]
\SetAlgoLined

\KwIn{temporal vectors $\vec b$, $\vec c$, $\vec d$, and $\vec e$}\

$\vec c' := \vec d\otimes \vec e$\tcp*[r]{temporal vector addition (requires two transitions), Figs.~\ref{fig:8sm_oper}(a,b)}
$\vec b' := \vec c \dashv \vec c'$\tcp*[r]{elementwise inhibit, Fig.~\ref{fig:8sm_oper}(d)}
$\vec a := \vec b \oplus \vec b'$\tcp*[r]{elementwise min, Fig.~\ref{fig:8sm_oper}(d)}\

\Return $\vec a$\;
\caption{Pseudocode for procedural computation of Eq.~\eqref{eq:toy_prb}}
\label{algo:toy}

\end{algorithm}
\end{figure}

This state machine allows arbitrary expressions such as
\begin{equation}
\label{eq:toy_prb}
    \vec a = \vec b \oplus' (\vec c \dashv (\vec d \otimes \vec e))
\end{equation}
to be calculated. The computation is performed by partitioning it into phases, with each phase implemented serially on the state machine of Fig~\ref{fig:8sm_oper}. By breaking the computation into discrete read-compute-store transitions of a state machine, we can represent the computation using a procedural algorithm, Algorithm~\ref{algo:toy}. 

We follow the regular order of arithmetic and perform the tropical multiplication first. Assume vector $\vec d$ and $\vec e$ reside in memories one and two. The $\otimes$ operation is shown in Fig.~\ref{fig:8sm_oper}(a),(b). The first phase selects the memory in the computation unit and applies a one-hot vector at the input of wavefront memory 1, initiating the computation. The memory places the vector $\vec d$ on the output data bus, which then passes it to the accumulator of the computation unit. The next step is shown in Fig.~\ref{fig:8sm_oper}(b), where memory 3 is setup to receive the output of the operation while being activated in write mode.  A one-hot vector is applied to the input of memory 2, playing the wavefront through the stored vector, and storing the resulting output in memory 3. This storage operation is indicated by the assignment operator $:=$ in the pseudocode. Tropical vector-matrix multiplication is a similar one-input operation and can be performed in a similar way, as shown in Fig.~\ref{fig:8sm_oper}(c). 

Two-operand operations such as elementwise inhibit, and tropical vector addition are all performed in the same way. Synchronized one-hot vectors are presented to the address input that causes output wavefronts to be triggered. These wavefronts enter the computational unit where circuits for the requested operations are multiplexed in, and the output is written to wavefront memory 3. This is all illustrated in Fig.~\ref{fig:8sm_oper}(d). In this way, one- and two-operand operations can be performed in a single state machine. Note that the computation is set up by control circuits not shown in the figure. These control circuits are the only circuits that are not temporal in nature, and are used to direct the flow of the computation in the system. These control circuits can be understood as the machine-level subroutines called by a ``tropical interpreter'' stepping through the lines of pseudocode in Algorithm~\ref{algo:toy}.

\subsection{A nontrivial example: DNA alignment}
\label{subsec:dna-align}

DNA alignment using a temporal instantiation of the Needleman-Wunsch algorithm was one of the first applications of race logic~\cite{madhavan2014race,madhavan20174}. In that work, the alignment matrix of the Needleman-Wunsch algorithm is physically laid out as a planar graph, and pure race logic operations define the scoring information at each node. Though the implementation in \cite{madhavan20174} is extremely fast and energy efficient, it suffers the disadvantage of requiring a dedicated ASIC. In this section we briefly sketch how Needleman-Wunsch might instead be implemented in a general-purpose tropical state machine like what we describe in the previous section. 

The Needleman-Wunsch algorithm finds the shortest path through a dynamically constructed score matrix. Each element of the score matrix $M_{ij}$ is constructed recursively as $M_{ij} = \min\{M_{i,j-1}+\sigma, M_{i-1,j} + \sigma, M_{i-1,j-1} + m\delta_{x_i,y_j}\}$, where $\sigma$ is the cost of a genetic insertion or deletion (an ``indel'') and $m$ is the cost of a single gene mutation.\footnote{The Kronecker delta $\delta_{ij}$ is defined as one when $i=j$ and zero otherwise.} This naturally has the structure of a tropical inner product, but the Kronecker delta function breaks the causality condition and so cannot be implemented in pure race logic.

To compute the Kronecker delta, we encode the set of four possible genes $\{\mathtt{G}, \mathtt{A}, \mathtt{T}, \mathtt{C}\}$ as temporal values $\{0,1,2,3\}$. We then use the coincidence function~\cite{smith2018space,am_isca_workshop2019} to determine equality of the temporally encoded gene values. Tropically the coincidence function is described as $\delta(t_1,t_2) = (t_1 \oplus t_2) \dashv (t_1 \oplus' t_2)$, which is equal to the inputs when they are the same,\footnote{The simple version presented in the text applies to only an idealized coincidence: the exact point where $t_1=t_2$. In practice~\cite{am_isca_workshop2019,smith2018space} a nonzero coincidence window can introduced via a tolerance $\epsilon$, by computing $[\epsilon\otimes(t_1\oplus t_2)]\dashv(t_1\oplus't_2)$.} and $\infty$ otherwise~\cite{am_isca_workshop2019}. The coincidence function could either be a primitive operation of the state machine or could be accomplished over multiple state transitions using $\oplus$, $\oplus'$, and $\dashv$; we assume the former circumstance. Binarization followed by projective storage of $\delta(x_i,y_j)$ would then save zero (tropical one) to memory when $x_i=y_j$ and $\infty$ (tropical zero) otherwise, resulting in a many-hot vector that indexes genewise equality. 

\begin{figure}
\removelatexerror
\begin{algorithm}[H]
\SetAlgoLined

\KwIn{gene sequences $\vec x, \vec y \in \{0,1,2,3\}^n$, indel cost $\sigma$, mismatch cost $m$}\

$\vec\mu^{(0)}$ := $[0]$\;
$\vec\mu^{(1)}$ := $[\sigma, \sigma]$\;
\

\tcp{Upper-left triangular part [$\mathrm{dim}(\vec\mu^{(k)})$ increasing]:}
\For{$k\leftarrow 2$ \KwTo $n$}{
  $\vec c' := \delta\left(\vec x_{1,\cdots,k-1}, \vec y_{k-1,\cdots,1}\right)$\tcp*[r]{$\text{mismatches}\rightarrow\infty$, $\text{matches} \rightarrow \{0,1,2,3\}$}
  $\vec c :\cong \text{binarize}(\vec c')$\tcp*[r]{$\text{mismatches}\rightsquigarrow\infty$, $\text{matches} \rightsquigarrow 0$}
  $\vec a := \sigma \otimes \vec\mu^{(k-1)}$\tcp*[r]{apply insertion/deletion (indel) cost $\sigma$}
  $\vec b := (m\oplus\vec c) \otimes \vec\mu^{(k-2)}$\tcp*[r]{apply mutation cost $m$ for mismatches}
  $\vec r :=  \vec a_{0,\cdots,k-2} \oplus \vec b \oplus \vec a_{1,\cdots,k-1}$\tcp*[r]{find least-cost local path (Eq.~\eqref{eq:nw-core})}
  $\vec \mu^{(k)} := \left[
    a_0,
    \vec r,
    a_{k-1}\right]$\tcp*[r]{append boundary conditions}
}\

\tcp{Lower-right triangular part [$\mathrm{dim}(\vec\mu^{(k)})$ decreasing]:}
\For{$k\leftarrow n+1$ \KwTo $2n$}{
  $\vec c' := \delta\left(\vec x_{k-n,\cdots,n}, \vec y_{n,\cdots,k-n}\right)$\tcp*[r]{$\text{mismatches}\rightarrow\infty$, $\text{matches} \rightarrow \{0,1,2,3\}$}
  $\vec c :\cong \text{binarize}(\vec c')$\tcp*[r]{$\text{mismatches}\rightsquigarrow\infty$, $\text{matches} \rightsquigarrow 0$}
  $\vec a := \sigma \otimes \vec\mu^{(k-1)}$\tcp*[r]{apply insertion/deletions (indel) cost $\sigma$}
  $\vec b := (m\oplus\vec c) \otimes \vec\mu^{(k-2)}$\tcp*[r]{apply mutation cost $m$ for mismatches}
  $\vec \mu^{(k)} := \vec a_{0,\cdots,2n-k} \oplus \vec b \oplus \vec a_{1,\cdots,2n-k+1}$\tcp*[r]{find least-cost local path (Eq.~\eqref{eq:nw-core})}
}\

\Return $\vec\mu^{(2n)}$\tcp*[r]{this is actually just a scalar: lowest possible alignment cost}
 \caption{Pseudocode for Needleman-Wunsch (forward pass only; computes optimal alignment cost)}
 \label{algo:needle}

\end{algorithm}
\end{figure}

To frame the Needleman-Wunch algorithm as a tropical vector problem, we exploit the independence of the skew-diagonals~\cite{lipton1985systolic}. We define $\vec\mu^{(k)}$ as the $k^\text{th}$ skew-diagonal vector of $M$, so that $\vec\mu^{(0)} = [M_{00}]$, $\vec\mu^{(1)}=[M_{10},M_{01}]$, and so on. The first and last elements of $\vec\mu^{(k)}$ are $k\sigma$ by construction for $k\leq n$, that is, until we hit the main skew diagonal. The defining equation for $M_{ij}$ is then given through $\vec\mu^{(k)}$ by 
\begin{equation}
\label{eq:nw-core}
    \mu^{(k)}_j = \left(\sigma\otimes\mu^{(k-1)}_j\right)
           \oplus \left(\left[m\oplus\delta_{x_j,y_{k-j}}\right]\otimes\mu^{(k-2)}_j\right)
           \oplus \left(\sigma\otimes\mu^{(k-1)}_{j+1}\right).
\end{equation}
The vectorized computation of this recursion relation is presented programmatically in Algorithm~\ref{algo:needle}. The right-hand side of each assignment is a pure race logic computation; the left-hand side represents a register address. As in Algorithm~\ref{algo:toy}, the assignment operator $\vec x := \vec y$ indicates storage of $\vec y$ to a temporal memory register represented by $\vec x$. The projective storage operator $\vec x :\cong \vec y$ assigns the tropical normalization $\vec y - \min \vec y$ to the vector register $\vec x$.

The interpreter required here is more complex than in Algorithm~\ref{algo:toy}. Though we could in principle implement the \textbf{for}-loops tropically by assigning $k := 1\otimes k$ and monitoring $n \dashv k$ and $2n \dashv k$, we are not aware of a way to elegantly perform subarray extraction using temporal signals as indices. We therefore imagine that $k$, as well as the array slicing operations, are managed digitally by the interpreter.

\section{Case study: Dijkstra's algorithm as implemented in a temporal state machine}
\label{sec:dijkstra}

%\begin{itemize}
%    \item step by step operation of each Dijkstra iteration. This is different from the phase of temporal computations. How one Dijkstra iteration is made of multiple temporal phases. 
%   \item describe state diagram and timing diagram with datapath for one Dijkstra's iteration. 
%   \item describe final backtrack operation. 
%end{itemize}

In Sec.~\ref{sec:temporal-state-machine} we demonstrate a simple model state machine, but it is too simple to utilize the graph traversal logic of tropical linear algebra that we describe in Sec.~\ref{subsec:graph_trop}. Though the Needleman-Wunsch machine in Sec.~\ref{subsec:dna-align} does perform graph traversal, it is restricted to a known, uniform progression through a highly regular planar graph. From the discussion of Sec.~\ref{subsec:graph_trop}, however, we know that general graph traversal should be accessible to a tropical state machine. In the present section, we discuss an implementation of Dijkstra's algorithm in a temporal state machine using the concepts developed in this paper. We will see that the core neighbor-search operation of Dijkstra's algorithm is naturally parallelized by the tropical VMM, leading to very high throughput in terms of graph edges traversed per unit time, and that the inhibit operation together with projective storage allow the embedding of important Boolean logic structures within the temporal framework. 

\subsection{Dijkstra's algorithm in race logic}

We assume that the reader is familiar with the classical implementation of Dijkstra's algorithm. In Algorithm~\ref{algo:dijkstra}, we map the operations of Dijkstra's algorithm into race logic, with each step as a single transition of a temporal state machine. Two trivial modifications simplify the race logic implementation. First, instead of tracking the known distances to each node, we mask out the distances of visited nodes with the value $\infty$. This vector of distances to unvisited nodes is $\vec d$ in the algorithm listing, and a tropically binarized record of which nodes have been visited is recorded in a vector $\vec v$. Second, instead of storing a parent vector directly, we define a parent matrix $\hat P$ as a collection of tropical column vectors where a finite entry $P_{ij}$ holds the distance from node $i$ to node $j$ along the current optimal path to $j$ from the source node $s$. We assume that the memristors in the VMM are already programmed to their correct values, meaning that the graph is already stored in the arithmetic unit. 

There are several apparent differences in how operations of the algorithm are performed in this  (tropical) linear algebra engine compared to a traditional programming language. There are, loosely speaking, two ``modes'' in which we use tropical vectors. First, there are true temporal wavefronts, such as $\vec e$ and $\vec d$, that represent variable distances measured throughout the graph. These flow through the data path of the algorithm. Second, there are indicator wavefronts, such as $\vec v$ and $\vec d^*$, with elements restricted to $0$ or $\infty$. These are used along the control paths of the algorithm to perform element lookup from data-carrying temporal wavefronts, modification of tropically binary records such as $\vec v$, and for index selection of the parent matrix. Projective storage plays a key role in these processes via binarization of one-hot vectors. Sometimes, quantities like $\vec n$ can play either of the above roles depending on context.

There are two primary constraints on this algorithm's application. First, because directed edge weights are encoded as temporal delays, negative edge weights are physically forbidden. 
Second, temporal vectors are limited to a finite dynamic range and resolution constrained by the technology in which they are implemented, and consecutive tropical multiplication could lead to dynamic range issues. To mitigate this dynamic range issue, we arrange the computation such that no more than one successive tropical multiplication occurs along a single datapath per state machine transition. Normalization of $\vec u$ at the end of each cycle shrinks the dynamic range as much as possible between VMMs.

\begin{figure}
\removelatexerror
\begin{algorithm}[H]
\SetAlgoLined
\KwIn{graph $G$, source node $s$}\

\tcp{Variable initializations}
$\vec d := \mathbf{0}_s$\tcp*[r]{distances to unvisited nodes (tropical one-hot labels source)}
$\vec v := \infty$\tcp*[r]{visited nodes (tropical zero vector)}
$\hat P := \infty$\tcp*[r]{parent matrix (tropical zero matrix)}
$\hat A$ := adjacency-matrix($G$)\tcp*[r]{adjacency matrix of the graph}\

\While{$\left(\bigoplus_j d_j < \infty\right)$}{
  $\vec n := \text{argmin}(\vec d)$\tcp*[r]{choose node to visit}\ %cwn,cn

  \tcp{Examine neighbors}
  $\vec e := \hat A \otimes \vec n$\tcp*[r]{VMM examine neighbors of current node}
  
  $\vec f := \vec d \dashv \vec e$\tcp*[r]{keep only newly-found shortest paths}\
  
  \tcp{Update records for the next iteration}
  $\vec v := \vec v \oplus \vec n$\tcp*[r]{record the current node as visited}
  $\vec d' := \vec d \oplus \vec f$\tcp*[r]{construct new record of shortest paths}
  $\vec d :\cong \vec v \dashv \vec d'$\tcp*[r]{update global unvisited distance vector}\
  
  \tcp{Parent vector update process}
  $\vec f^* :\cong \text{binarize}(\vec f)$\tcp*[r]{vector indices of found nodes}
  $\hat P := \vec f^* \dashv \hat P$\tcp*[r]{delete row data of previously recorded parents for found nodes}
  $\vec P_{\vec n} := \vec f$\tcp*[r]{record in column $\vec n$ distances $\vec f$ from $\vec n$ to the found nodes}
}\

\Return $\hat P$\tcp*[r]{adjacency matrix of the minimal spanning (from $s$) subgraph of $G$}
\caption{Pseudocode for Temporal Dijkstra's Algorithm}
\label{algo:dijkstra}
\end{algorithm}
\end{figure}

The algorithm initializes by setting the vector $\vec d$ of known distances to unvisited nodes to a tropical one-hot $\mathbf{0}_s$ labeling the source node $s$. The vector $\vec v$ labeling visited nodes, as well as the parent matrix $\hat P$ keeping track of the minimal spanning tree through the graph, have all elements set to $\infty$. We assume the weighted adjacency matrix $\hat A$ of the desired graph has been programmed to a VMM unit before the algorithm begins. This is a one-time cost that can be amortized over frequent reuse of the array. The algorithm then begins by cycling the state machine through the main loop.

In each iteration, we check to see if any unvisited nodes have are available for exploration by evaluating the minimum element of $\vec d$. The algorithm terminates if this operation returns $\infty$, which indicates that all nodes have either been visited or are unreachable. Taking the argmin (Sec.~\ref{subsec:rl_other}) of $\vec d$ nominally gives us a vector $d_j \otimes\mathbf{0}_j$ where $j$ is the index of a node along a shortest path (of those so far explored) from the source and $\mathbf{0}_j$ is the tropical one-hot labeling index $j$. This can be thought of as a one-hot vector with a magnitude $d_j$.  However, we will see that $d_j$ is always zero by construction, so $\text{argmin}(\vec d)$ is just a one-hot. We store this one-hot to the vector register $\vec n$.

The next step is to examine the directed edges to the neighbors of node $\vec n$. We use $\vec n$ as the input to a temporal VMM operation with $\hat A$, which performs a parallel traversal to all neighbors. The result of this exploration is stored in $\vec e$. This vector may contain shorter paths to the neighbors nodes, via node $\vec n$, than what had been previously found. Such shorter paths would manifest as elements of $\vec e$ with smaller values than their corresponding elements in $\vec d$. Those specific nodes can be extracted by taking an elementwise inhibit of $\vec e$ by $\vec d$; the resulting updated distance vector is stored as $\vec d'$. We also note that $\vec n$ has been visited, and should not be visited again, by imposing the zero of $\vec n$ onto $\vec v$ and saving it in memory.

If the dynamic range of our memory were boundless, we could perform this operation repeatedly and determine the final distance vector of the algorithm. But because we are dynamic-range-limited,\footnote{Note that even if our memory were not range-limited, we must still choose a dynamic-range cutoff at which we assign finite time values to $\infty$; otherwise, a circuit that outputs $\infty$ as a valid return value could never halt. In practice, though, memory is the limiting factor. However the finite delay representing $\infty$ is chosen, it informs the effective clock frequency of the state machine.} we have to ensure that accumulation in the distance vector is minimized. We do this via projective storage of $\vec d'$ into $\vec d$. We also inhibit $\vec d'$ by $\vec v$ before storage to ensure that no nodes we have already visited are candidates for exploration in the next iteration; this also ensures $\text{argmin}(\vec d)$ will be a magnitude-free one-hot on the next cycle.  This shifts the temporal origin for the entirety of the next iteration into the perspective of $\text{argmin}(\vec v \dashv \vec d')$; all temporal values in the new $\vec d$ are now expressed \emph{relative} to the stopwatch of an observer at the argmin node. 

After completing neighbor-exploration, we update the parent matrix. The newly found nodes in $\vec f$ are the ones whose parents need to be updated. A binarized version $\vec f^*$ of $\vec f$ is used to inhibit rows of the parent matrix corresponding to the new paths in $f$, erasing these nodes' now-outdated parent data. This operation is performed row-by-row, requiring $N$ state machine transitions to complete. The new parent is then added to the parent matrix; $\vec n$ is used to enable the appropriate column of $\hat P$ for writing. Vector $\vec f$ is then written to this column. 

Throughout this algorithm, we require dynamical indexing of memory addresses based on past results of the temporal computation. Recall that Needlemen-Wunsch algorithm required significantly nontrivial subarray selection operations in Algorithm~\ref{algo:needle}. We claimed in Sec.~\ref{subsec:dna-align} that these would likely needed to be handled digitally. Those index selections can be statically determined at compile time, so they could merely be part of the elaborated bytecode controlling the state machine: there is no need for data to translate back and forth between temporal and digital domains in order to execute Algorithm~\ref{algo:needle}. In Algorithm~\ref{algo:dijkstra}, index selections of the parent matrix are dynamically determined at runtime, and cannot be statically embedded in the digital controller around the state machine. But the one-hot nature of the indexing operations offer natural interfaces to the crossbar architecture, so, again no digital intermediary is required to perform address lookup.

% Having made it to the end of this section, I feel like the shortening of
% some of these variable names was unnecessary. There's plenty of room in the
% algorithm environment to use "distances" instead of "dist" and "currentNode"
% instead of "cn". maybe we should just be explicit and help the reader not
% have to learn short variable names.

\section{Results}
\label{sec:performance-results}

We start the evaluation of this temporal state machine by describing the assumptions in its design and the simulation framework we use. To understand the scaling of this architecture, we create models for temporal memories and the tropical operations required by its design.  This process allows us to understand the trade-offs of our design space and make predictions about optimization targets. 

In order to achieve realistic first order performance estimates for this temporal state machine, we designed and simulated each of the components of our system using commercial very-large-scale integrated circuit (VLSI) design tools with experimentally validated nanodevice models~\cite{jiang2014verilog,chen2015compact}.  These devices exhibit voltage and current ranges typical of other memristors fabricated by a variety of groups~\cite{chakrabarti2017multiply,adam20163,yu2018neuro,xia2019memristive}. Though the voltage needed to read these devices can be low ($\approx 200$~mV), the voltages needed to write them can be as high as 2~V to 3~V, which puts a lower limit on the technology node we can use. To secure enough voltage headroom for changing device states, we use the 180~nm Silterra\footnote{Certain commercial processes are identified in this paper to foster understanding. Such identification does not imply recommendation or endorsement by the National Institute of Standards and Technology, nor does it imply that the processes identified are necessarily the best available for the purpose.} process with a $V_{dd} = 1.8$~V. Though this may not offer the most energy-efficient results, it does provide an understanding of the general set of trade-offs involved in building realistic temporal state machines. We provide a description of scaling to lower technology nodes by referring to the scaling laws presented in~\cite{stillmaker2017scaling}. A discussion on resistive switching technologies at deep-sub-micron nodes has been reported in section~\ref{subsec:tech_consderations_disc}.

In this work, the temporal memory is memristive, as is discussed in~\cite{madhavan2020storing}. The core cell is composed of a 1T1R structure with supporting circuits that allow temporal read and write operations. The temporal read operation is performed by down-shifting the input voltage level from $1.8$~V to $600$~mV before applying it to the 1T1R array, so that the device state is unaffected. This causes the output voltage to have a maximum value of 600~mV which needs to be up-shifted to $1.8$~V for compatibility with other functional blocks, all of which work at $V_{dd}$. The write path of the memory includes circuits for two different write modes, the conventional and normalized forms described previously. Both these operations require similar circuits with an input first-arrival detector charging the source line and level-shifting circuits to the appropriate write voltages, causing the quasilinear state write described in~\cite{madhavan2020storing}. 

\begin{figure*}
    \centering
    \includegraphics[width=\linewidth]{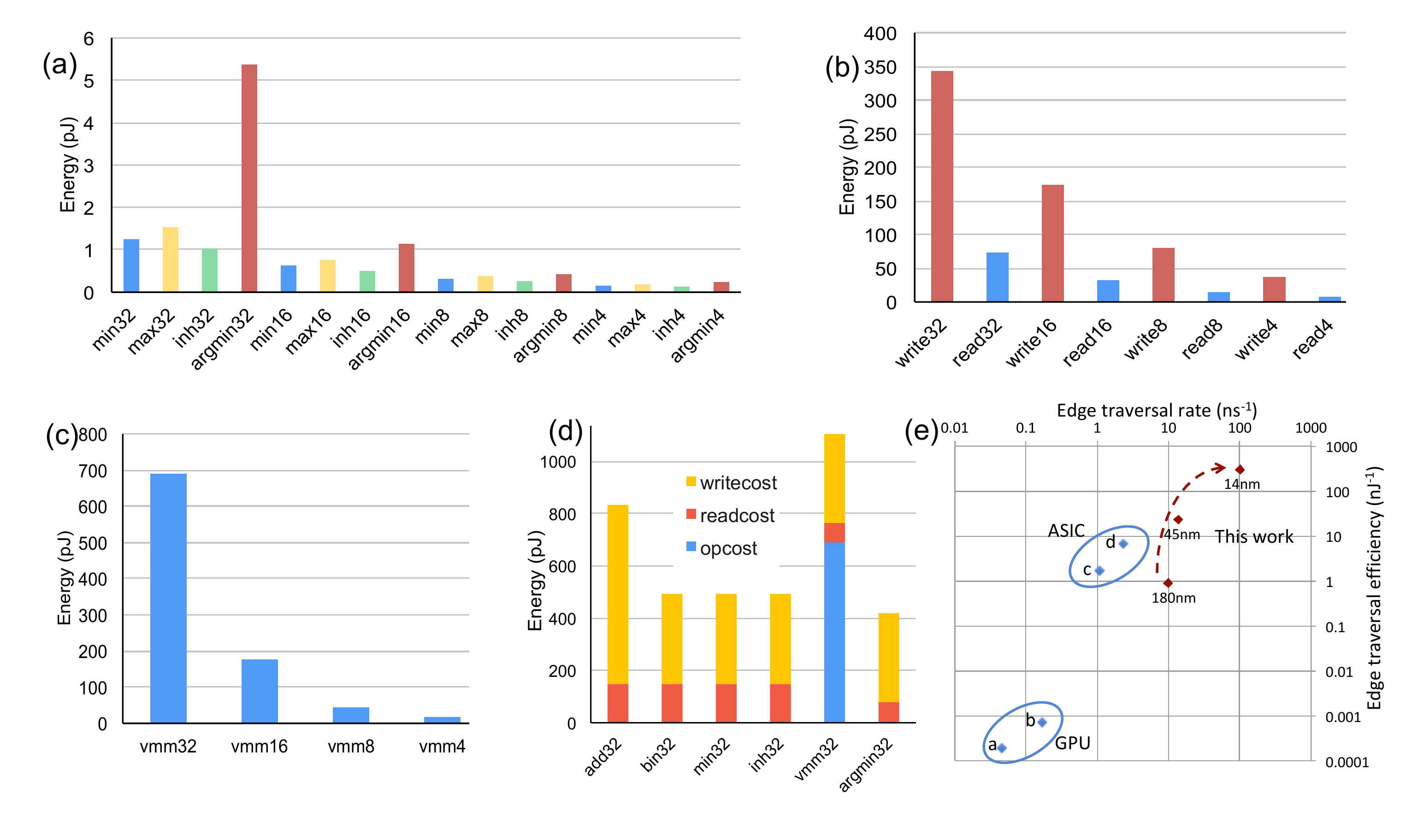}
    \caption{Energy results for various operations: Panel (a) shows the energy costs for vector operations of various array sizes. Panel (b) shows the energy costs for read and write operations using the memristor temporal memory, while panel(c) shows the energy costs of the VMM operations. Panel (d) compiles these energies and presents the energy cost of single and multi-operand operations in a temporal state machine of size $32\times32$. Panel (e) compares the energy cost of such a $32\times32$ kernel with that of state-of-the-art application specific integrated circuit (ASIC) and graphical processing unit (GPU) designs. It shows simulation results from our 180~nm process as well scaling to more advanced nodes following the procedure described in~\cite{stillmaker2017scaling}.}
    \label{fig:9_final_results}
\end{figure*}

The read and write energy costs for various $N\times N$ array sizes ranging from $N=4$ to $N=32$ are shown in Fig.~\ref{fig:9_final_results}. The energy scales superlinearly with array size due to growth in support circuitry size that scales with $N$, the input driver needs to be scaled up for larger array sizes; for the write case, larger array sizes require first-arrival circuitry with more inputs. From the figures, we see that the read cost is approx 2~pJ per line while the write cost is around 10~pJ per line. This $5\times$ factor between read and write energies drives many of the tradeoff considerations in designs.

The most computationally intensive pure race logic function is the tropical VMM, which implements a single-step all-to-all graph traversal. Such an operation naturally scales as  $N^2$, which can be seen in Fig.~\ref{fig:9_final_results}(b). On average, this system ends up costing $\approx 700$~fJ per cell, so a $32\times32$ grid consumes $\approx 700$~pJ of energy. The large energy cost of this operation arises from the conservative design strategy we employed. In order to make sure that the \textsc{or} pulldown network functions properly, we have to ensure that the time constants of the pulldown dynamics are not determined by the CMOS---that is, we have to ensure that it switches quicker than the resolution of our temporal code. The low read voltage causes the pulldown transistor to discharge too slowly, causing multiple nodes pulling down the same source line and leading to functional incorrectness of tropical addition. In order to overcome this issue, we add level-shifters to each cell to boost the input voltage from $V_\text{read}$ to $V_{dd}$. These provide the necessary overdrive for correct operation. 

Other pure race logic functions such $\oplus = \min$, $\oplus' = \max$, and $\dashv\; = \text{inhibit}$---as well as compound functions such as argmin and binarize---are implemented with conventional CMOS gates and hence have a minimal energy cost for this process node. For example, for a 32 channel elementwise-$\oplus$, the energy cost is approximately 1~pJ. This is negligible compared to temporal read, write, and VMM operations. The argmin operation has the largest energy cost among the combinatorial gates, since the first arriving input has to turn of all of the other channels and must therefore drive circuits with a larger output capacitance. The energy cost of each of these operations with respect to the problem size is shown in Fig.~\ref{fig:9_final_results}(a).

The energy numbers for the key phases of Algorithm~\ref{algo:dijkstra} for a problem size of 32 are shown in Fig.~\ref{fig:9_final_results}(d). Every operation incurs a single write cost, by virtue of the output that has to be written in to the memory cells, except for $\otimes$, which incurs two writes [Fig.~\ref{fig:8sm_oper}(a,b)]. Read costs, on the other hand, depend on the number of operands.  Single operand operations such as argmin require a single read, while binary operations such as $\oplus$, $\oplus'$, and $\dashv$ incur the twice the read costs. Energy costs for all operations except tropical VMM are dominated by reads and writes. This is as a result of the simplicity of the primitive operations which are essentially made of simple Boolean primitives.

\section{Comparisons and Technical considerations}
\label{sec:comp_tech_considerations}

% In this section, we contrast the presented architecture with other approaches and discuss a methodology to make a fair comparison among emerging paradigms. First we present a discussion of various technical challenges faced when building such temporal circuits, as well as a discussion on future research directions to make temporal computing systems more practical. Then we address technical considerations relevant to future iterations of our temporal state machine architecture.

\subsection{Comparison with previous work}
\label{subsec:rl_comp}
Graph processing is a well studied problem in computing and a variety of solutions have been proposed for it at various scales~\cite{gui2019survey}. Processing of real world graphs---which can contain hundreds of thousands of nodes and millions of edges---combines both software and hardware frameworks, employing everything from central processing units (CPUs), field programmable gate arrays (FPGAs)~\cite{zhang2018degree,zhou2017accelerating}, and graphics processing units (GPUs)~\cite{fu2014mapgraph,wang2016gunrock} to application specific integrated circuits (ASICs)~\cite{ham2016graphicionado,yan2019alleviating} and processing-in-memory (PIM) solutions~\cite{song2018graphr,zhou2019gram,challapalle2020gaas}. Graph operations are known to have a high communication-to-computation ratio, as the cost of memory movement sometimes accounts for upwards of 90~\% of total energy expenditures. The simple temporal architecture presented in Sec.~\ref{sec:dijkstra} is not developed adequately to sensibly compare it to such highly developed systems optimized for much larger graphs.  Another impediment to fair comparison, echoed by the authors of~\cite{gui2019survey}, is that much of the extant literature reports relative improvements against other implementations without providing absolute numbers for comparison. This makes comparison with PIM implementations especially difficult. The purpose of this work is to demonstrate the viability of temporal computing as an general approach using a well-studied example, not to compete with the best graph processing engines.

Therefore, we take the following approach. We do not compare against performant CPU and FPGA approaches that leverage 3D-stacked high-bandwidth memory (HBM) or hybrid memory cube (HMC), since these approaches rely on the memory management system for their performance advantages. GPU and ASIC approaches with domain-specific kernel implementations amortize the costs of these memory accesses much more effectively and are more popular. For example, Map Graph~\cite{fu2014mapgraph} and Gunrock~\cite{wang2016gunrock} are examples of GPU-based graph analytics packages commonly used as a baseline when reporting performance. More recently, domain-specific accelerators have emerged that have custom datapaths, scheduling strategies, scratchpad memory, and other techniques specifically designed to alleviate the irregularities associated with graph analytics. The literature on these approaches effectively reports the memory versus processing costs~\cite{ham2016graphicionado,yan2019alleviating}, allowing us to compare just the performance of our kernel with the performance of other state-of-the-art graph kernels. Under this analysis, one could imagine swapping in temporal state machines for existing subgraph kernels and measuring changes in overall performance metrics.

The metric widely used to make speed comparisons is the edge traversal rate, commonly reported as giga-edge traversals per second (GETS) in the literature. For energy efficiency, we speak of edges traversed per unit energy---giga-edge traversals per joule (GETJ) in the literature. Figure \ref{fig:9_final_results}(e) shows the performance comparison of this work against GPU and ASIC approaches.  A single $32 \times 32$ kernel in a 180~nm technology node has an edge traversal rate of 10~ns$^{-1}$ (10 GETS) and the energy efficiency is about 1~nJ$^{-1}$ (1~GETJ), which compares favorably with the state of the art. Using scaling projections from~\cite{stillmaker2017scaling}, we estimate that a single kernel can theoretically surpass state-of-the-art kernel performance. When scaled up to larger $N\times N$ array sizes, such as $N=128$ or $N=256$ (not an uncommon core size for memristor crossbars), we can expect massive performance improvements. Note that the state of the art for graph processing engines when energy is of no concern is on the order of 100s of GETS, which our analysis indicates to be feasible for temporal designs. 

Independent but parallel work on graph problems is being undertaken by the neuromorphic computing community. Dikjstra's algorithm has been studied by researchers in neuromorphic computing as a benchmark application for the field~\cite{davies2019benchmarks,hamilton2019spike,aimone2019dynamic}. State-of-the-art industrial research spiking neural network platforms~\cite{davies2018loihi} use Dijkstra's algorithm to establish performance metrics for their systems. The Tennessee neuromorphic project uses single-source shortest path computation to demonstrate their spiking neuromorphic chips~\cite{schuman2019shortest}. The energy per spike costs have been detailed in Ref.~\cite{schuman2019shortest}; when an operation equivalent to the tropical VMM is implemented, it costs approximately 2.5~nJ in a 65~nm process. By comparison, combining both the memory and VMM primitives, race logic performs the same operation for 1~nJ in a 180~nm process. 
%We believe that the conventional CMOS approaches should be a north star, guiding the performance that emerging paradigms, when sufficiently developed, should be able to surpass.

\subsection{Technical considerations}
\label{subsec:tech_consderations_disc}

%In this section we discuss some frequently asked questions regarding the choices made in the design of the architecture/

\subsubsection{Scaling from 180~nm to newer technology nodes.}
Previous work with race logic has demonstrated that most of the energy expended in race logic architectures is spent in the distribution of timing information, such as in clock trees or analog voltages~\cite{madhavan20174}. In order to get an energy advantage over those approaches, the present work relies on novel technologies such as memristors to locally generate a programmable delay, which has the advantage that the energy cost is limited by the capacitor. Hence we are limited by today's memristor technology, which requires large write voltages (1.2~V to 6~V). This requires that we use a relatively old technology node. The development of memristor technology is being driven toward the goal of CMOS compatibility at advanced technology nodes, which require lower read and write voltages~\cite{choudhary2019low}. Companies are exploring low write voltage resistive random access memory (ReRAM) and embedding it into 22~nm fin field-effect transistor (FinFET) stacks~\cite{jain201913,golonzka2019non}. 

Maturing technology has great promise for the designs proposed in this work. As CMOS transistors become smaller, the area, energy, and speed all improve. For example, when moving from a 180~nm CMOS to 14~nm FinFET, using a fan-out-of-4 inverter as a benchmark, the area, energy, and latency numbers improve by 100$\times$, 190$\times$ and 19$\times$, respectively~\cite{stillmaker2017scaling}. As memristor technologies become compatible with lower voltages, the energy of the read and write operations are expected to decrease. The write energy, determined by the voltage and current needed to alter the memristive state, changes less than the read energy, which follows the inverter characteristics. The scaling performance of race logic systems is easy to estimate since the spatial nature of the information flow ensures that the architectures in various technology nodes all have similar design and activity factors. We expect the dynamic energy cost to follow the energy trend of the inverter as described in Ref.~\cite{stillmaker2017scaling}. Though latency and area are determined by other factors such as memory dynamic range and functional correctness, the overall advantage in energy-delay-product from scaling to a lower node could be as high as three orders of magnitude. 

\subsubsection{Memristor device non-idealities.}

There are a variety of device non-idealities that affect the design. First, the dynamic range of the technology is limited. Prior work with memristors has demonstrated precision as high as 6~bits to 7~bits~\cite{xia2019memristive,alibart2012high}, but this was accomplished with a carefully programmed write-verify process. With the memristor model used in this paper, we can extract up to 5~bits of precision. Practical implementations have even lower precision. One way to increase precision is to use extra wires to encode higher precision bits as done for Boolean logic. A similar idea has been proposed in Refs.~\cite{chen201919,kim20149}.

Another major impediment to smooth operation in our circuits is the linearity requirement of the memristor write process. A truly linear write would increase the dynamic range of our operations and ensure a clear mapping between the read and write processes. This linearity requirement has been a major topic of research for the neuromorphic VMM community with large implications on the hardware training of large scale neural networks~\cite{xia2019memristive,sun2019impact,burr2015experimental}. Considerable effort has been dedicated to this effort. Various groups show highly linear behaviour by operating in the high conductance regime with proper compliance control~\cite{li2018analogue}, exploring new materials~\cite{chandrasekaran2019improving}, and using three terminal lithium devices~\cite{fuller2017li}. An alternate approach utilizes highly linear trench-capacitor based storage~\cite{li2018capacitor}. Recently, a temporal magnetic memory has been proposed which exhibits linear dynamics~\cite{vakili2020temporal}. This proposal re-purposes magnetic configurations in racetracks such as domain walls or skyrmions to encode temporal information spatially within the race track.

Finally, issues such as noise, variability, drift, and mismatch will be ultimately responsible for determining the actual dynamic range that can be extracted from such nanodevices~\cite{xia2019memristive}. The advantage of using a charge-based approach to computation is that the memristors can be used in their high-conductance, low-variability regime, while still maintaining low read energy costs. Building a noise model that describes how variability accumulates in such systems is beyond the scope of this text, but will be an important future work. The energy benefits of such the analog approach come at the cost of error; effectively bounding this tradeoff is a crucial theoretical problem.

\subsubsection{Future design considerations.}

Temporal computation leads to unconventional architectures that come with their own design constraints. The cost of primitive operations (aside from the VMM) in temporal computing is cheap compared to memory access operations. This points to utilizing strategies that amortize the cost of memory accesses over multiple feed-forward operations. Future systems would greatly benefit by performing many such operations in a single phase. In Algorithm~\ref{algo:dijkstra}, for instance, neither $\vec e$ nor $\vec d'$ need to be stored in memory. A sophisticated compiler could detect optimally long compositions of pure race logic functions and only use memory where invariance or causality need to be broken. Though such a state machine would need additional control logic with separate clock and dummy lines, the energy savings accrued by this sort of optimization would be significant.

As higher level algorithms become more clearly expressible, an important question would be,  what kind of complexity of operations would we want in our designs? Similar to the discussion of reduced versus complex instruction set computers (RISC vs. CISC), a design with simpler fundamental primitives could be more flexible, but might sacrifice performance. An example of that can be seen in the parent matrix update of Algorithm~\ref{algo:dijkstra}. A 2D update array similar to the VMM could amortize the cost of $N$ extra operations, and hence save on $N$ memory reads and writes, in just a single operation. Hence a more complex operation would have smaller energy and delay, which would be very favorable---at the cost of specialized circuitry. The sensibility of such tradeoffs is an open question that needs to be addressed.

Finally, we must consider the question of dynamic range. Approaches to extend the dynamic range of such memories by using a binary-like encoding has been proposed by previous authors~\cite{kim20149}. These techniques may be required to expand dynamic ranges to be compatible with future designs.

\section{Conclusion}
\label{sec:conclusion}

The utility of temporal computation in solving problems expressible by dynamic programming has been widely noted. Though the first race logic work was proposed as a hardware acceleration for dynamic programming algorithms, it was constrained in its design: a limited topology and a feed-forward memoryless structure. Only the length of the shortest path was reported, with extra circuitry nominally required to report the path itself. Since then, other designs with state-of-the-art performance have been proposed, but they similarly suffer from an \textit{ad hoc} design approach. 

In this work we attempt to make the first steps at generalizability of temporal computing. We provide a generalizeable data-path and a mathematical algebra, expanding the logical framework of race logic . This leads to novel circuit designs that are informed by higher level algorithmic requirements. The properties of abstraction and composability offered by the mathematical framework coupled with native storage from the temporal memory lend themselves to generalization. We design a state machine that can carry out both specialized and general graph algorithms, such as Needleman-Wunsch and Dijkstra's algorithm, respectively. The potential for general purpose graph accelerators built on temporal computing motivates further exploration of temporal state machines.

\section{Acknowledgments}

Authors thank Brian Hoskins, Mark Anders, Jabez McClelland, Melika Payvand, George Tzimpragos, and Tim Sherwood for helpful discussions. 
Advait Madhavan acknowledges support under the Cooperative Research Agreement Award No.~70NANB14H209, through the University of Maryland.

\bibliographystyle{ACM-Reference-Format}
\bibliography{references.bib}

\end{document}